\documentclass[11pt,a4paper,oneside]{article}
\usepackage[top=3cm, bottom=3cm, left=2cm, right=2cm]{geometry}
\usepackage[english]{babel}
\usepackage[utf8x]{inputenc}
\usepackage[T1]{fontenc}

\usepackage{amsthm,amsmath,amssymb,amsfonts,mathrsfs,dsfont,mathtools}

\usepackage{bm}
\usepackage{graphicx}
\usepackage{booktabs,subcaption}
\usepackage[all]{nowidow}
\usepackage{microtype}

\usepackage{setspace}
\usepackage[ruled,norelsize,vlined]{algorithm2e}
\usepackage[dvipsnames]{xcolor}
\usepackage{tikz} 
\usepackage[colorlinks=true, allcolors=blue]{hyperref}

\usepackage[sf]{titlesec}
\usepackage{sectsty}
\allsectionsfont{\centering \normalfont\scshape} 

\usepackage[round]{natbib}

\usepackage{authblk}
\title{Enriched Pitman--Yor processes}
\date{}
\author[1]{Tommaso Rigon}
\author[2]{Bruno Scarpa}
\author[3]{Sonia Petrone}
\affil[1]{Department of Economics, Management and Statistics, University of Milano--Bicocca, 20126 Milano, Italy}
\affil[2]{Department of Statistical Sciences, Universit\`a degli studi di Padova, Padova, Italy}
\affil[3]{Department of Decision Sciences, Bocconi University, Via Roentgen 1, Milano, Italy}

\newtheorem{theorem}{Theorem}
\newtheorem{corollary}{Corollary}

\newtheorem{proposition}{Proposition}

\theoremstyle{definition}
\newtheorem{definition}{Definition}
\newtheorem{remark}{Remark}

\makeatletter
\def\env@cases{%
  \let\@ifnextchar\new@ifnextchar
  \left\lbrace
  \def\arraystretch{0.8}%
  \array{@{}l@{\quad}l@{}}}
\makeatother

\newcommand \dd  { \,\textup d}   

\newcommand{\iidsim}{\overset{\textup{iid}}{\sim}}
\newcommand{\tp}{\tilde{p}}

\begin{document}
\maketitle

\begin{abstract}
In Bayesian nonparametrics there exists a rich variety of discrete priors, including the Dirichlet process and its generalizations, which are nowadays well-established tools. Despite the remarkable advances, few proposals are tailored for modeling observations lying on product spaces, such as~$\mathds{R}^p$. 
Indeed, for multivariate random measures, most available priors lack flexibility and do not allow for separate partition structures among the spaces. We introduce a discrete nonparametric prior, termed enriched Pitman--Yor process (\textsc{epy}), aimed at addressing these issues. Theoretical properties of this novel prior are extensively investigated. We discuss its formal link with the enriched Dirichlet process and normalized random measures, we describe a square-breaking representation and we obtain closed-form expressions for the posterior law and the involved urn schemes. 
In second place, we show that several existing approaches, including Dirichlet processes with a spike and slab base measure and mixture of mixtures models, implicitly rely on special cases of the \textsc{epy}, which therefore constitutes a unified probabilistic framework for many Bayesian nonparametric priors. Interestingly, our unifying formulation will allow us to naturally extend these models while preserving their analytical tractability. As an illustration, we employ the \textsc{epy} for a species sampling problem in ecology and for functional clustering in an e-commerce application. 
\end{abstract}
\section{Introduction}\label{sec:intro}

The Dirichlet process (\textsc{dp}) of \citet{Ferguson1973} is a widely employed nonparametric \emph{discrete} prior which arguably stands because of its analytical tractability. Despite its popularity, the \textsc{dp} may be a restrictive modeling choice as it depends on a single parameter controlling both the variability and the random partition it induces. To overcome these limitations several proposals have been made, including the Pitman--Yor  process (\textsc{py}) discussed in \citet{Perman1992}, \citet{Pitman1997}, and the very general classes of Gibbs-type priors \citep{DeBlasi2015}, normalized random measures with independent increments \citep{ReLP2003, Lijoi2007}, and species sampling models \citep{Pitman1996}. These extensions offer a richer modeling framework while preserving the tractability of the \textsc{dp}. One may refer to \citet{Lijoi2010} for an overview.

Unfortunately, none of the aforementioned priors has been specifically designed to model exchangeable observations lying on a product space, such as~$\mathds{R}^p$.  More precisely, let $\mathds{X}$ and $\mathds{Y}$ be two complete and separable Polish spaces and let $\left((X_n,Y_n)\right)_{n \ge 1}$ be an infinite sequence of exchangeable random elements taking values in the product space $\mathds{X} \times \mathds{Y}$, for example $\mathds{R}^p=\mathds{R}^k \times \mathds{R}^{p-k}$. Then, de Finetti theorem guarantees that conditionally on a random probability measure $\tilde{p}$ the random elements $(X_n,Y_n)$ are independent and identically distributed (iid), namely
\begin{equation}\label{joint_iid}
\begin{aligned}
(X_n, Y_n) \mid \tilde{p} & \overset{\textup{iid}}{\sim} \tilde{p}, \qquad n \ge 1,\\
\tilde{p} &\sim \mathcal{Q},
\end{aligned}
\end{equation}
where $\mathcal{Q}$ is the probability law of the random $\tp$, namely the \emph{prior} distribution. In a first motivating application, the random variables $X_n$ and $Y_n$ represent families and species of trees in the Amazonian basin, respectively. Given a collection of observations, we are then interested in predicting the number of novel families and species that one would get in a future sample. However, neither the \textsc{dp}, the \textsc{py}, or other species sampling models are suitable priors $\mathcal{Q}$. Indeed, these choices would associate the discovery of a new species with that of a new family, which is an unrealistic assumption. Instead, we seek priors inducing nested mechanisms, so that the discovery of a new family of trees corresponds to that of a new species, but not vice versa. In the second motivating application, we aim at forming groups of functional observations, each representing the number of searches on a website carried out over time by different customers. The random variables $X_n$ and $Y_n$ represent latent features and latent functions, respectively, which induce a nested and more interpretable clustering mechanism that will turn useful for market segmentation. 

A more suitable prior for the joint random probability measure $\tilde{p}$ in model~\eqref{joint_iid} is the enriched Dirichlet process (\textsc{edp}) of \citet{Wade2011}. The \textsc{edp} allows for finer control of the dependence structure between the $X_n$ and $Y_n$ and leads to the desired \emph{nested clustering}. The \textsc{edp} broadens the principles of enriched conjugate priors for natural exponential families \citep{Consonni2001} to the nonparametric setting since it extends the construction of the enriched Dirichlet distribution of \citet{Connor1969} to random probability measures; refer to \citet{Wade2011} for further details. These appealing features have been exploited among others in \citet{Wade2014, Gadd2019} for Bayesian nonparametric regression models, in \citet{Roy2018} for causal inference with missing covariates, and in \citet{Zeldow2021} for functional clustering of longitudinal data. 

Here we move away from the original derivation of the \textsc{edp}, considering constructions beyond conjugacy. The \textsc{edp} inherits also some of the drawbacks shared by all Dirichlet-based priors, which indeed motivates our extension. For instance, the probability of observing a new species in the urn scheme of the \textsc{dp} \citep[i.e.][]{Blackwell1973} solely depends on the sample size and not on the previously observed values. As remarked by \citet{Lijoi2007b}, this simplifying assumption is particularly problematic in species sampling problems, because the posterior probability of discovering a new species does not depend on the data. In addition, this feature of the \textsc{dp} leads to a logarithmic growth of the number of clusters and to a lack of robustness with respect to miscalibrated prior choices, which might be undesirable in several applied contexts; see e.g. \citet{Lijoi2007} and \citet{DeBlasi2015}. 

We address these issues by proposing a novel discrete prior law $\mathcal{Q}$, that builds upon the \textsc{edp} and the \textsc{py} process. By combining their appealing properties, the proposed \emph{enriched Pitman--Yor} process~{(\textsc{epy})} leads to different rates for the number of clusters and more robust Bayesian estimators for species sampling models. Importantly, improved flexibility is attained while preserving analytical and computational tractability. We obtain a simple urn scheme, a tractable posterior characterization, and a so-called square-breaking representation. In addition, we show that the \textsc{epy} can be defined by normalizing a suitable random measure. This alternative definition parallels the construction of \citet{ReLP2003} and has important theoretical implications. 

Beside their key role for the modeling of observations in product spaces, \textsc{epy} priors are provably useful also in other settings. Specifically, consider an exchangeable sequence $(Y_n)_{n \ge 1}$ taking values in $\mathds{Y}$ and let
\begin{equation}\label{marginal_y}
\begin{aligned}
Y_n \mid \tilde{p}_Y &\overset{\textup{iid}}{\sim} \tilde{p}_Y, \qquad n \ge 1, \\
\tilde{p}_Y &\sim \mathcal{Q}_Y.
\end{aligned}
\end{equation}
Hence, one can consider a \emph{marginal} \textsc{epy} process as the prior $\mathcal{Q}_Y$, which is defined as 
\begin{equation}\label{marginalEPY}
\tilde{p}_Y(B) = \tilde{p}(\mathds{X} \times B), \qquad \tilde{p} \sim \mathcal{Q},
\end{equation}
for any Borel set $B \subseteq \mathds{Y}$, where the prior law $\mathcal{Q}$ is an \textsc{epy} process on $\mathds{X} \times \mathds{Y}$ and the space $\mathds{X}$ should be interpreted as a latent dimension that induces enriched specifications.  It will be shown that the marginal \textsc{epy} $\tilde{p}_Y$ is an infinite mixture of \textsc{py} processes, which is arguably much more flexible than a single \textsc{py} process. 

We show that model~\eqref{marginal_y} includes a rich variety of prior proposals in the literature as a special case. To the best of our knowledge, their connection with enriched processes has not been previously emphasized. For example, a specific marginal \textsc{epy} process has been implicitly studied in \citet{Scarpa2014} and \citet{Rigon2019} for the analysis of functional data. Dirichlet processes with spike and slab base measures \citep[e.g.][]{MacLehose2007,Dunson2008b,Guindani2009}, or general atomic contaminations \citep{Scarpa2009}, are actually special cases of a marginal \textsc{edp}. A mixture of finite-dimensional \textsc{dp}s has been employed in \citet{Malsiner2017} and \citet{Rigon2019} to perform model-based clustering, while convex combinations of \textsc{dp}s have been considered by \citet{Muller2004}, \citet{Lijoi2014b} to induce dependence across groups of random variables. These models are also strongly linked to the marginal \textsc{epy}. Through the paper, we will point out the connections between the \textsc{epy} and the aforementioned methods, aiming at providing a unified probabilistic framework for these classes of processes. 

The paper is organized as follows. In Section~\ref{sec:2} we introduce the \textsc{epy} process and we discuss its fundamental probabilistic characterizations, including the square-breaking construction. In Section~\ref{sec:posterior} we discuss an enriched urn scheme and posterior representations. In Section~\ref{sec:methods} we illustrate the marginal \textsc{epy} process and its connection with several existing approaches, and we propose numerous extensions. In Section~\ref{sec:app1} we employ the \textsc{epy}  to estimate the number of unobserved species in the Amazonian tree flora.  Finally, in Section~\ref{sec:app2} we employ the \textsc{epy} in a mixture model for functional clustering, illustrating its usefulness for market segmentation. Concluding remarks are given in Section~\ref{discussion}. All the proofs are collected in the Supplementary Material.

\section{The enriched Pitman--Yor process}\label{sec:2}

The \textsc{epy} is built upon the \textsc{dp} and the \textsc{py} processes, of which we provide a concise overview that is also useful to set the notation. One can refer to \citet{Lijoi2010} and \citet{DeBlasi2015} for more structured reviews. The \textsc{epy} process is then defined in Section~\ref{sec:epy}, together with the so-called square-breaking representation. 

\subsection{Background material}\label{sec:background}

The \textsc{py} process is a probability law on a random discrete distribution, that can be defined through the so-called stick-breaking construction. Let $P$ be a probability measure on $\mathds{X} \times \mathds{Y}$ and let $(\nu_j)_{j\ge 1}$ be a sequence of independent Beta random variables with $\nu_j \sim \textsc{beta}(1-\sigma,\alpha + j\sigma)$, where either $\sigma\in[0,1)$ and $\alpha> - \sigma$, or $\sigma = -\alpha/H$ and $\alpha > 0$ for some integer $H \in \{2,3,\dots \}$. A discrete random probability measure $\tilde{p}$ follows a \textsc{py} process with parameters $(\sigma, \alpha P)$, written $\tilde{p} \sim \textsc{py}(\sigma, \alpha P)$, if 
\begin{equation}\label{py_def}
\tilde{p}(\cdot) = \sum_{h = 1}^\infty \xi_h\:\delta_{(\phi_h, \theta_h)}(\cdot), \qquad (\phi_h,\theta_h) \overset{\textup{iid}}{\sim} P,
\end{equation}
with $\xi_h=\nu_h\prod_{j=1}^{h-1}(1-\nu_j)$ for $h\ge 1$, where we agree that $\xi_1 = \nu_1$. The parameter $\sigma$ is often called discount or stable  parameter whereas $\alpha$ is termed total mass or precision. If $\sigma = 0$, then $\tilde{p}$ in equation~\eqref{py_def} defines a \textsc{dp} and we write $\tilde{p} \sim \textsc{dp}(\alpha P)$. If instead $\sigma = -\alpha/H$ and $\alpha > 0$, then the stick-breaking construction is degenerate because $\nu_H = 1$, implying that $\sum_{h=1}^H\xi_h =1$ for a finite integer $H < \infty$. This special case of \textsc{py}, called Dirichlet multinomial process, or Fisher process, admits the following alternative representation
\begin{equation*}
\tilde{p}(\cdot) \overset{\textup{d}}{=} \sum_{h=1}^H W_h\:\delta_{(\phi_h, \theta_h)}(\cdot), \qquad (W_1,\dots,W_H)\sim \textsc{dirichlet}(\alpha/H,\dots,\alpha/H),
\end{equation*}
where $(\phi_h,\theta_h) \overset{\textup{iid}}{\sim} P$ and with $\overset{\textup{d}}{=}$ denoting the equality in distribution. Thus, when $\sigma < 0$, the \textsc{py} reduces to a finite-dimensional discrete prior law having symmetric Dirichlet weights. One may refer to the Appendix A.1 of \citet{Pitman1997} for such a distributional equivalence. 

Although the \textsc{dp} is a special case of~\eqref{py_def}, throughout the paper we will make extensive use of an alternative construction based on completely random measures. This approach is somewhat less straightforward compared to~\eqref{py_def}, but it has important theoretical implications. Broadly speaking, a \textsc{dp} can be obtained as the normalization of a Gamma process \citep{Ferguson1973}, which in turn can be represented as $\tilde{\mu}(\cdot) = \sum_{h=1}^\infty J_h \:\delta_{(\tilde{\phi}_h,\tilde{\theta}_h)}(\cdot)$, where $J_1 \ge J_2 \ge \dots$ is a collection of ordered positive random jumps whose distribution is given in \citet{Ferguson1972}, that are independent on the random locations $(\tilde{\phi}_h,\tilde{\theta}_h) \overset{\textup{iid}}{\sim} P$. The law of a Gamma random measure $\tilde{\mu}$ is uniquely characterized by its Laplace functional, namely
\begin{equation}\label{laplace_Gamma}
\mathds{E}\left\{e^{- \tilde{\mu}(f)}\right\} = \exp\left\{ - \alpha \int_{\mathds{X} \times \mathds{Y} }\log\left\{1 + f(x,y)\right\}  P(\dd x, \dd y) \right\},
\end{equation}
with $\alpha >0$ and for any positive and measurable function $f : \mathds{X} \times \mathds{Y}  \rightarrow \mathds{R}^+$ such that $\tilde{\mu}(f) = \int_{\mathds{X} \times \mathds{Y}}f(x,y) \tilde{\mu}(\dd x, \dd y) < \infty$ almost surely. We write $\tilde{\mu} \sim \textsc{gap}(\alpha P)$.  The \textsc{dp} is then defined as the normalization of $\tilde{\mu}$, that is, if $\tilde{\mu} \sim \textsc{gap}(\alpha P)$ then $\tilde{p}(\cdot) = \tilde{\mu}(\cdot) / \tilde{\mu}(\mathds{X} \times \mathds{Y}) \sim \textsc{dp}(\alpha P)$.

\subsection{Definition and alternative representations}\label{sec:epy}
Informally, the \textsc{epy} process is obtained by integrating a collection of independent Pitman--Yor processes over a Dirichlet process. The definition has the same rationale that underlies the construction of enriched conjugate priors for natural exponential families \citep{Consonni2001}, which is extended to nonparametric settings in the \textsc{edp} and, to some extent and for univariate random measures, in neutral to the right processes \citep{Doksum1974}. 
Roughly speaking, a multivariate distribution is decomposed in terms of the marginal of and the conditional distributions. Then, an enriched prior law on the parameters of the joint distribution is obtained by assigning independent conjugate priors on the parameters of the marginal and on those of the conditional distributions. In the nonparametric case, the construction is more delicate as the distributions involved are {\em random} probability measures. 

\begin{definition}\label{def2} Let $P$ be a probability measure on the product space $\mathds{X} \times \mathds{Y}$, with $P(A \times B) = \int_A P_{Y \mid X}(B \mid x) P_X(\dd x)$ for any Borel sets $A \subseteq \mathds{X}$ and $B \subseteq \mathds{Y}$. Moreover, let $\alpha > 0$ and $\sigma(x)$, $\beta(x)$ be functions such that either $\sigma(x) \in [0,1)$ and $\beta(x) > - \sigma(x)$ or $\sigma(x) = -\beta(x)/H(x)$ and $\beta(x) > 0$, where  $H(x) \in \{2,3,\dots\}$. Define a random probability measure $\tilde{p}_X$ on $\mathds{X}$ and a family of random probability measures $\tilde{p}_{Y \mid X}(\cdot \mid x)$ on $\mathds{Y}$, for $x \in \mathds{X}$, such that
\begin{equation*}
\tilde{p}_X \sim \textsc{dp}(\alpha P_X),\qquad \tilde{p}_{Y \mid X}(\cdot \mid x) \overset{\textup{ind}}{\sim} \textsc{py}\{\sigma(x), \beta(x) P_{Y \mid X}(\cdot \mid x) \}, \qquad x \in \mathds{X},
\end{equation*}
independently among themselves. Then the random probability measure $\tilde{p}$ on the product space $\mathds{X} \times \mathds{Y}$, defined as
\begin{equation} \label{eq:map}
\tilde{p}(A \times B) = \int_A \tilde{p}_{Y \mid X}(B \mid x) \tilde{p}_X(\dd x),  \qquad A \subseteq \mathds{X}, \quad B \subseteq \mathds{Y},
\end{equation}
is said to be distributed as an \emph{enriched Pitman--Yor} process (\textsc{epy}) with parameters $\alpha P_X, \sigma(x)$ and $\beta(x) P_{Y \mid X}$.  We will write $\tilde{p} \sim \textsc{epy}(\alpha P_X, \sigma(x), \beta(x) P_{Y \mid X})$.
\end{definition}
Hence, the \textsc{edp} of \citet{Wade2011} is a special case of the \textsc{epy} if $\sigma(x) = 0$ for any $x \in \mathds{X}$. Ensuring that an \textsc{epy} process is a well defined stochastic process, namely that it provides a law of the random joint probability measure $\tilde{p}$, is a quite subtle measure-theoretic issue,  involving random conditional distributions. However, this can be shown along the steps given in \citet{Wade2011} for the \textsc{edp}, which in turn are based on the results of \citet{Ram2006}. Indeed, the only conditions there required are that the conditionals $\tilde{p}_{Y \mid X}( \cdot \mid x)$ are independent across $x \in \mathds{X}$, the marginal $\tp_X$ is a.s. discrete and $\tp_X$ and $\tp_{Y \mid X}$ are independent among themselves; all these properties hold for the \textsc{epy} by construction. A detailed proof is provided in the Supplementary Material for completeness.   

Note that the \textsc{dp} on the product space $\mathds{X} \times \mathds{Y}$ is a limiting case of the \textsc{epy}, occurring when $\beta(x) \rightarrow - \sigma(x)$ for any $x \in \mathds{X}$. Indeed, at the limit, each conditional law reduces to a point mass, i.e. $\tilde{p}_{Y \mid X}(B\mid x) = \delta_{\theta_1(x)}(B)$, with $\theta_1(x) \sim P_{Y \mid X}(\cdot;x)$, therefore 
\begin{equation*}
\tilde{p}(A \times B) = \sum_{h=1}^\infty \xi_h \: \delta_{\phi_h}(A) \delta_{\theta_1(\phi_h)}(B) = \sum_{h=1}^\infty \xi_h \: \delta_{(\phi_h,\theta_h)}(A \times B), \qquad (\phi_h,\theta_h) \overset{\textup{iid}}{\sim} P, 
\end{equation*}
as each $\beta(x) \rightarrow -\sigma(x)$, where $(\xi_h)_{h \ge 1}$ are the stick-breaking weights of $\tilde{p}_X$. Summarizing, an \textsc{epy} process $\tilde{p}$ on $\mathds{X} \times \mathds{Y}$ reduces to a $\textsc{dp}(\alpha P)$, as $\beta(x) \rightarrow - \sigma(x)$ for any $x \in \mathds{X}$.  In addition, note that the baseline measures $P_X$, $P_{Y \mid X}$ and $P$ can be interpreted as ``prior guesses'' for the distribution of the observations, because one has
\begin{equation*}
\mathds{E}\{\tilde{p}_X(A)\} = P_X(A), \quad \mathds{E}\{\tilde{p}_{Y \mid X}(B \mid x)\} = P_{Y \mid X}(B \mid x), \quad \mathds{E}\{\tilde{p}(A \times B)\} = P(A \times B),
\end{equation*}
for any $x \in \mathds{X}$ and Borel sets $A \subseteq \mathds{X}$, $B \subseteq \mathds{Y}$, recalling that $P(A \times B) = \int_A P_{Y \mid X}(B \mid x)P_X(\dd x)$. 

The \textsc{epy} can be alternatively defined through a square-breaking representation. Such an equivalent definition, presented in the next proposition, is important especially for computational reasons, as one might truncate the involved series to approximate the infinite-dimensional process; see e.g. \citet{Ishwaran2001,Scarpa2014}. In addition, it emphasizes that the \textsc{epy} is a discrete random probability measure. 

\begin{proposition}\label{stbr} Let the quantities $\alpha P_X, \sigma(x), \beta(x) P_{Y \mid X}$ be as in Definition~\ref{def2} and let $\tilde{p}$ be a random probability measure on $\mathds{X} \times \mathds{Y}$ such that $\tilde{p} \sim \textsc{epy}(\alpha P_X, \sigma(x), \beta(x) P_{Y \mid X})$. Then
\begin{equation*}
\tilde{p}(A \times B) =\sum_{\ell =1}^\infty \sum_{h=1}^\infty \xi_\ell\:\pi_h(\phi_\ell)\:\delta_{\phi_\ell}(A)\:\delta_{\theta_h(\phi_\ell)}(B), \qquad A\subseteq \mathds{X},\quad B\subseteq \mathds{Y},
\end{equation*}
where $\xi_\ell=\nu_\ell\prod_{j=1}^{\ell-1}(1-\nu_j)$ and $\pi_h(x)=\eta_h(x)\prod_{j=1}^{h-1}\{1-\eta_j(x)\}$ for any $\ell \ge 1$ and $h \ge 1$, with
\begin{equation*}
\begin{aligned}
\nu_j &\overset{\textup{iid}}{\sim} \textsc{beta}(1,\alpha), \qquad &&\phi_\ell \overset{\textup{iid}}{\sim} P_X, \\
\eta_j(x) &\overset{\textup{ind}}{\sim} \textsc{beta}\{1-\sigma(x),\beta(x) + j\sigma(x)\}, \qquad &&\theta_h(x) \overset{\textup{iid}}{\sim} P_{Y \mid X}(\cdot\mid x),
\end{aligned}
\end{equation*}
independently among themselves for any $j \ge 1$, $\ell \ge 1$, $h \ge 1$ and $x \in \mathds{X}$.
\end{proposition}

When $\sigma(x) = 0$, one recovers the square-breaking construction of the \textsc{edp} given in \citet{Wade2011}. Moreover, if the discount parameter $\sigma(x)$ is strictly negative for any $x \in \mathds{X}$, then the square-breaking representation simplifies, because in this case the  conditional laws $\tilde{p}_{Y \mid X}$ of the \textsc{epy} process are finite-dimensional; refer to Section~\ref{sec:background}.  

Paralleling the construction of the \textsc{dp}, we present a third alternative definition of the \textsc{epy} process, through the normalization of a random measure that we call \emph{Gamma and Pitman--Yor} process (\textsc{ga-py}). 

\begin{definition}\label{def1} Let the quantities $\alpha P_X, \sigma(x), \beta(x) P_{Y \mid X}$ be as in Definition~\ref{def2}. Define a random probability measure $\tilde{\mu}_X$ on $\mathds{X}$ and a family of random probability measures $\tilde{p}_{Y \mid X}(\cdot \mid x)$ on $\mathds{Y}$, for $x \in \mathds{X}$, such that $\tilde{\mu}_X \sim \textsc{gap}(\alpha P_X)$ and $\tilde{p}_{Y \mid X}(\cdot \mid x) \overset{\textup{ind}}{\sim} \textsc{py}\{\sigma(x), \beta(x) P_{Y \mid X}(\cdot \mid x) \}$, independently among themselves. Then the random probability measure $\tilde{\mu}$ on the product space $\mathds{X} \times \mathds{Y}$, defined as
\begin{equation*}
\tilde{\mu}(A \times B) = \int_A \tilde{p}_{Y \mid X}(B \mid x) \tilde{\mu}_X(\dd x),  \qquad A \subseteq \mathds{X}, \quad B \subseteq \mathds{Y},
\end{equation*}
is said to be distributed as a \emph{Gamma and Pitman--Yor} process (\textsc{ga-py}) with parameters $\alpha P_X, \sigma(x)$ and $\beta(x) P_{Y \mid X}$.  We will write $\tilde{\mu} \sim \textsc{ga-py}(\alpha P_X, \sigma(x), \beta(x) P_{Y \mid X})$.
\end{definition}

Let $\tilde{\mu}$ be a random measure with $\tilde{\mu} \sim \textsc{ga-py}(\alpha P_X,\sigma(x),  \beta(x) P_{Y \mid X})$. Then, the random probability measure $\tilde{p}$ on the product space $\mathds{X} \times \mathds{Y}$
\begin{equation*}
\tilde{p}(\cdot) = \frac{\tilde{\mu}(\cdot)}{\tilde{\mu}(\mathds{X} \times \mathds{Y})},
\end{equation*}
is distributed according to an \emph{enriched Pitman--Yor} process (\textsc{epy}) with parameters $\alpha P_X, \sigma(x)$ and $\beta(x) P_{Y \mid X}$. The normalizing constant $\tilde{\mu}(\mathds{X} \times \mathds{Y})$ in the above definition is a positive random variable such that $\tilde{\mu}(\mathds{X} \times \mathds{Y})= \int_{\mathds{X}}\tilde{p}_{Y \mid X}(\mathds{Y}\mid x)\tilde{\mu}_X(\dd x) = \tilde{\mu}_X(\mathds{X})$ almost surely (a.s.). Hence, for any Borel sets $A \subseteq \mathds{X}$ and $B \subseteq \mathds{Y}$, an \textsc{epy} process can be written as follows
\begin{equation}\label{epy_def2}
\begin{aligned}
\tilde{p}(A \times B) &= \int_A \tilde{p}_{Y \mid X}(B\mid x) \frac{\tilde{\mu}_X}{\tilde{\mu}_X(\mathds{X})}(\dd x) = \int_A \tilde{p}_{Y \mid X}(B\mid x) \tilde{p}_X(\dd x),
\end{aligned}
\end{equation}
where $\tilde{p}_X(\cdot) = \tilde{\mu}_X(\cdot)/\tilde{\mu}_X(\mathds{X}) \sim \textsc{dp}(\alpha P_X)$.

\section{Predictive rule and posterior law}\label{sec:posterior}

We now turn to the investigation of the conditional properties of the \textsc{epy}. Consider the exchangeable random sequence $\left((X_n,Y_n)\right)_{n\ge 1}$ such that
\begin{equation*}
(X_n,Y_n) \mid \tilde{p} \overset{\textup{iid}}{\sim}\tilde{p}, \qquad  \tilde{p} \sim \textsc{epy}(\alpha P_X, \sigma(x), \beta(x) P_{Y \mid X}), \qquad n \ge 1.
\end{equation*}
Note that we can sample each $(X_n,Y_n)$ through a two-step mechanism. Specifically, $X_n$ is first drawn from the marginal distribution $\tilde{p}_X$ and then, given $X_n = x$, each $Y_n$ is obtained from the conditional distribution $\tilde{p}_{Y \mid X}(\cdot \mid x)$; more precisely,
\begin{equation}\label{enriched_sampling}
\begin{aligned}
X_n \mid \tilde{p}_X  \overset{\textup{iid}}{\sim}\tilde{p}_X, \qquad Y_n \mid X_n = x, \ \tilde{p}_{Y \mid X}(\cdot \mid x) \overset{\textup{ind}}{\sim}\tilde{p}_{Y \mid X}(\cdot \mid x), \qquad n \ge 1.
\end{aligned}
\end{equation}
Recall that, if $\tilde{p}$ is an \textsc{epy}, the marginal law $\tilde{p}_X$ is a $\textsc{dp}(\alpha P_X)$ whereas each conditional $\tilde{p}_{Y \mid X}$ is a $\textsc{py}(\sigma(x), \beta(x)P_{Y \mid X})$, independently among themselves and on $\tilde{p}_X$ for any $x \in \mathds{X}$. Our results will easily follow from~\eqref{enriched_sampling} and well-known properties of \textsc{dp} and \textsc{py} processes. To facilitate their derivation, we shall assume that each conditional baseline measure $P_{Y \mid X}(\cdot \mid x)$ is a.s. \emph{diffuse}, that is, it does not have a discrete component. In contrast, the marginal baseline measure $P_X$ may have atoms or even being discrete. Indeed, in the \textsc{py} case the assumption of a diffuse baseline measure is essential to avoid more complicate probabilistic calculations  \citep{Camerlenghi2019}.

\subsection{Enriched urn scheme}\label{sec:urns}

We discuss an enriched urn scheme that extends the predictive mechanism of the \textsc{edp} given in \citet{Wade2011}. By ``urn scheme'' we broadly mean an urn-like predictive rule, such as the one described in \citet{Blackwell1973}. For the \textsc{epy}, the urn scheme provided in the next theorem sheds light on the underlying random partition and highlights the importance of the additional set of parameters $\sigma(x)$, which indeed allows for a much finer calibration of the random partition compared to the \textsc{edp}.

The a.s. discreteness of the marginal law $\tilde{p}_X$ implies that, in model~\eqref{enriched_sampling}, there will be ties in the realization $\bm{x}^{(n)} = (x_1,\dots,x_n)$ of the random variables $\bm{X}^{(n)} = (X_1,\dots,X_n)$, with positive probability. Let $(x_1^*,\dots,x_{k_x}^*)$ denote the $k_x$ distinct values within $\bm{x}^{(n)}$, in order of appearance, with associated frequencies $(n_1,\dots,n_{k_x})$, so that $n_1 + \cdots + n_{k_x} = n$. In the sampling mechanism~\eqref{enriched_sampling}, the random variables $(Y_{1r},\dots,Y_{{n_r}r})$ associated to the $r$th distinct value $x_r^*$ are conditionally iid draws from the discrete distribution $\tilde{p}_{Y \mid X}(\cdot\mid x_r^*)$. Thus, with positive probability there will be further ties in each sample $(Y_{1r},\dots,Y_{{n_r}r})$, with distinct values $(y_{1r}^*,\dots,y_{{k_r}r}^*)$, in order of appearance, and with frequencies $(n_{1r},\dots,n_{k_r r})$, so that $n_{1r} + \cdots + n_{k_r r} = n_r$. Hence, the number of distinct values $k_y = k_1 + \cdots + k_{k_x}$ within a realization $\bm{y}^{(n)} = (y_1,\dots,y_n)$ of $\bm{Y}^{(n)} = (Y_1,\dots,Y_n)$ is such that $k_x \le k_y$.  This two-step stochastic mechanism can be described through an urn scheme, as illustrated by the next theorem.

\begin{theorem}\label{predictive} Suppose $\left((X_n,Y_n)\right)_{n \ge 1}$ is an exchangeable sequence as in equation~\eqref{joint_iid}, with $\tilde{p} \sim \textsc{epy}$ $(\alpha P_X, \sigma(x), \beta(x) P_{Y \mid X})$. Moreover, suppose that for each $x \in \mathds{X}$ the probability measure $P_{Y \mid X}(\cdot \mid x)$ is a.s. diffuse. Then, $(X_1,Y_1) \sim P$ and for any $n \ge 1$
\begin{equation}\label{pred1}
\begin{aligned}
X_{n+1} \mid \bm{X}^{(n)} = \bm{x}^{(n)} &\sim \frac{\alpha}{\alpha + n}P_X + \frac{1}{\alpha+n} \sum_{r=1}^{k_x} n_r\:\delta_{x^*_r},
\end{aligned}
\end{equation}
where $(x_1^*,\dots,x_{k_x}^*)$ are the $k_x$ distinct values within $\bm{x}^{(n)}$ with frequencies $(n_1,\dots,n_{k_x})$. Moreover,  for any $r = 1,\dots,k_x$ and $n \ge 1$
\begin{equation}\label{pred2}
\begin{aligned}
&Y_{n+1} \mid \bm{X}^{(n)} = \bm{x}^{(n)}, \bm{Y}^{(n)} = \bm{y}^{(n)}, X_{n+1} = x 
\\
&\qquad\sim \frac{\beta_r + k_r \sigma_r}{\beta_r + n_r}P_{Y \mid X}(\cdot \mid x_r^*) + \frac{1}{\beta_r + n_r}\sum_{j=1}^{k_r}(n_{jr} - \sigma_r) \:\delta_{y_{jr}^*}, && \textup{ if } x = x_r^*,\\
&\qquad\sim P_{Y \mid X}(\cdot \mid x), && \textup{ if } x \not\in\{x_1^*,\dots,x_{k_x}^*\}
,
\end{aligned}
\end{equation}
with $\beta_r = \beta(x_r^*)$ and $\sigma_r = \sigma(x_r^*)$, where $(y_{1r},\dots,y_{{n_r}r})$ are the values of $\bm{y}^{(n)}$ associated to $x_r^*$, whereas $(y_{1r}^*,\dots,y_{{k_r}r}^*)$ are the corresponding $k_r$ distinct values, with frequencies $(n_{1r},\dots,n_{k_r r})$, for $r = 1,\dots,k_x$. 
\end{theorem}

\begin{remark}\label{rem1} The system of predictive laws in the above theorem uniquely characterizes the \textsc{epy} process; refer to the Supplementary Material for details. 
\end{remark}

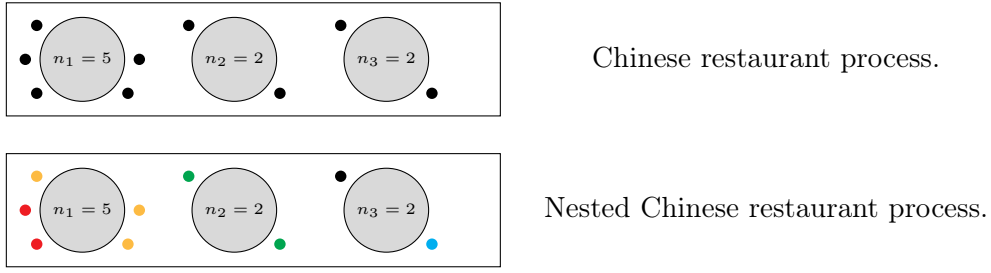
\begin{figure}[tb]
\centering
\begin{tikzpicture}

  \draw (-4, 0) rectangle (2.5, 1.5);
  \node at (-3,0.75) [circle, draw=black!100, fill=black!15] {\tiny{$n_1 = 5$}};
  \fill (-3.6, 0.3) circle (0.075cm);
  \fill (-3.75, 0.75) circle (0.075cm);
  \fill (-2.25, 0.75) circle (0.075cm);
  \fill (-2.4, 0.3) circle (0.075cm);
  \fill (-3.6, 1.2) circle (0.075cm);
  
  \node at (-1,0.75) [circle, draw=black!100, fill=black!15] {\tiny{$n_2 = 2$}};
  \fill (-1.6, 1.2) circle (0.075cm);
  \fill (-0.4, 0.3) circle (0.075cm);

  \node at (1,0.75) [circle, draw=black!100, fill=black!15] {\tiny{$n_3 = 2$}};
  \fill (1.6, 0.3) circle (0.075cm);
  \fill (0.4, 1.2) circle (0.075cm);
  
  \draw (6, 0.75) node {Chinese restaurant process.};

  \draw (-4, -2) rectangle (2.5, -0.5);
  \node at (-3,-1.25) [circle, draw=black!100, fill=black!15] {\tiny{$n_1 = 5$}};
  \fill[Red] (-3.6, -1.7) circle (0.075cm);
  \fill[Red] (-3.75, -1.25) circle (0.075cm);
  \fill[Dandelion] (-2.25, -1.25) circle (0.075cm);
  \fill[Dandelion] (-2.4, -1.7) circle (0.075cm);
  \fill[Dandelion] (-3.6, -0.8) circle (0.075cm);
  
  \node at (-1,-1.25) [circle, draw=black!100, fill=black!15] {\tiny{$n_2 = 2$}};
  \fill[Green] (-1.6, -0.8) circle (0.075cm);
  \fill[Green] (-0.4, -1.7) circle (0.075cm);
  
    \node at (1,-1.25) [circle, draw=black!100, fill=black!15] {\tiny{$n_3 = 2$}};
  \fill (0.4, -0.8) circle (0.075cm);
  \fill[Cyan] (1.6, -1.7) circle (0.075cm);

  \draw (6, -1.25) node {Nested Chinese restaurant process.};
\end{tikzpicture}
\caption{Classical and nested Chinese restaurant metaphor: circles represent tables, bullets represent customers and  colors represent dishes. The number of customers for each table are: $n_1 = 5$, $n_2 = 2$ and $n_3=2$, for a total of $n = \sum_{r=1}^3 n_r = 9$ customers. In the nested metaphor, customers are sub-partitioned according to dishes, with frequencies $n_{11} = 3$ (yellow), $n_{21} = 2$ (red), $n_{12} = 2$ (green), $n_{13} = 1$ (black), $n_{23} = 1$ (blue). \label{metaphor_pic}}
\end{figure}

The two-stage random partition implied by the predictive rule can be described in terms of a nested Chinese restaurant (\textsc{crp}) metaphor, illustrated in Figure~\ref{metaphor_pic}, which extends the one of \citet{Wade2014}. Let us assume that $P$ is diffuse almost surely. Suppose there exists a restaurant with a potentially infinite number of tables, representing the $X_n$, which serves a potentially infinite number of dishes, representing the $Y_n$. A first customer seats in one of the tables and selects a dish. For $n \ge 1$, the $(n+1)$th customer may either sit in one of the occupied  tables, say the $r$th, with probability $n_r / (\alpha + n)$ for $r=1,\dots,k_x$, or she can seat in a new one with probability $\alpha/(\alpha + n)$. If a new table is chosen, she will get a new dish. Otherwise, she may either select a new dish with probability $(\beta_r + k_r \sigma_r) / (\beta_r + n_r)$ or choose one of dishes previously served at her table, say the $j$th, with probability $(n_{jr} - \sigma_r)/(\beta_r + n_r)$, for $j=1,\dots,k_r$.  In comparison, the classical Chinese restaurant process only partitions customers in tables, disregarding the dish of choice. Such a nested \textsc{crp} describes the random partition implied by the \textsc{epy}. Moreover, if we label the tables with iid draws from $P_X$, and the dishes at the $x$th table with iid draws from $P_{Y \mid X}(\cdot \mid x)$, then one obtains the enriched P\'olya sequence $\left((X_n, Y_n)\right)_{n \ge 1}$ defined by the predictive rule \eqref{pred1}-\eqref{pred2}.

If $\sigma(x) = 0$ for any $x \in \mathds{X}$, then the sequence $((X_n, Y_n))_{n \geq 1}$ with predictive rule as in Theorem~\ref{predictive} corresponds to the enriched P\'olya sequence described in \citet{Wade2011}. Under the hypothesis of Corollary~\ref{cor2}, $((X_n, Y_n))_{n \geq 1}$ is a P\'olya sequence \citep{Blackwell1973}.  Moreover, this also occurs when $\beta(x) \rightarrow -\sigma(x)$ for any $x \in \mathds{X}$, consistently with the discussion in Section~\ref{sec:epy}. As in the \textsc{edp}, the precision parameter $\alpha$ and the function $\beta(x)$ regulate the number of distinct values within $\bm{X}^{(n)}$ and $\bm{Y}^{(n)}$. However, the additional function $\sigma(x)$ controls the asymptotic clustering behavior of each subsequence $(Y_{1r},\dots,Y_{{n_r}r})$, given $\bm{X}^{(n)} = \bm{x}^{(n)}$, that is, the growth rate of the number of clusters of $(Y_{1r},\dots,Y_{{n_r}r})$. In addition, the discount parameter $\sigma(x)$ allows to regulate the variance of the within-group number of clusters $k_r$, leading to more robust specifications. It is hence clear that the \textsc{epy} allows for a much  greater flexibility compared to the \textsc{edp}. We refer to \citet{Lijoi2007} and \citet{DeBlasi2015} for an extensive discussion about the role of the discount parameter~$\sigma(x)$ and its usefulness both for species sampling and mixture models.

For example, positive values of $\sigma(x) \in (0,1)$ lead to a within-group polynomial growth rate of the number of distinct values, which is much faster than the logarithmic rate of the \textsc{dp}, occurring when $\sigma(x) = 0$. Conversely, if the discount parameter is negative, i.e. $\sigma(x) = - \beta(x)/ H(x)$  with $\beta(x) >0$, then number of clusters is bounded by $H(x)$. Indeed, in this case for any $n \ge 1$
\begin{equation*}
\begin{aligned}
&Y_{n+1} \mid \bm{X}^{(n)} = \bm{x}^{(n)}, \bm{Y}^{(n)} = \bm{y}^{(n)}, X_{n+1} = x \\
&\qquad  \sim \left(1 - \frac{k_r}{H_r}\right)\frac{\beta_r}{\beta_r + n_r}P_{Y \mid X}(\cdot \mid x_r^*) + \frac{1}{\beta_r + n_r}\sum_{j=1}^{k_r}(n_{jr} + \beta_r / H_r) \:\delta_{y_{jr}^*}, \quad  x = x_r^*,
\end{aligned}
\end{equation*}
with $\beta_r = \beta(x_r^*), \sigma_r = \sigma(x_r^*)$ and $H_r = H(x_r^*)$. The above equation highlights that the within-group number of clusters cannot be greater than $H_r$, a feature which has been provably useful in several applied contexts, including the application of Section~\ref{sec:app2}.

\subsection{Posterior distribution}\label{subsec:posterior}

We now derive the posterior law of the random probability measure $\tilde{p}$. The \textsc{epy} process is not conjugate but the corresponding posterior is nonetheless analytically tractable. Recall that, by definition, $\tp (A \times B) = \int_A \tilde{p}_{Y \mid X}(B\mid x) \tilde{p}_X(\dd x)$, for $A \subseteq \mathds{X}$  and $B\subseteq \mathds{Y}$. Therefore, its posterior distribution may be obtained, at least in principle, from the posterior laws of the random marginal distribution $\tilde{p}_X$  and of the conditionals $\tilde{p}_{Y \mid X}(\cdot \mid x)$. Those posterior distributions are obtained in the following theorem.

\begin{theorem}\label{posterior} Let $(X_i,Y_i) \mid \tilde{p} \overset{\textup{iid}}{\sim} \tilde{p}$ for $i=1,\dots,n$, with $\tilde{p} \sim \textsc{epy}(\alpha P_X, \sigma(x), \beta(x) P_{Y \mid X})$, and suppose that each conditional probability measures $P_{Y \mid X}(\cdot \mid x)$ is a.s. diffuse. Then, under the notation of Theorem~\ref{predictive}, one has
\begin{equation*}
\tilde{p}_X \mid \bm{X}^{(n)} = \bm{x}^{(n)},\bm{Y}^{(n)} = \bm{y}^{(n)} \sim\textsc{dp}\left(\alpha P_X + \sum_{r=1}^{k_x} n_r\:\delta_{x^*_r} \right).
\end{equation*}
Moreover, for any $x \in \mathds{X}$ one has
\begin{equation*}
\begin{aligned}
\tilde{p}_{Y \mid X}(\cdot \mid x) \mid \bm{X}^{(n)} = \bm{x}^{(n)}, \bm{Y}^{(n)}= \bm{y}^{(n)} &\overset{\textup{d}}{=} W_{0r} \:\tilde{p}_{Y \mid X}^*(\cdot \mid x_r^*) + \sum_{j=1}^{k_r} W_{jr} \:\delta_{y_{jr}^*}(\cdot),  \ x = x_r^*, \\
\tilde{p}_{Y \mid X}(\cdot \mid x) \mid \bm{X}^{(n)} = \bm{x}^{(n)}, \bm{Y}^{(n)}= \bm{y}^{(n)} &\sim \textsc{py}\{\sigma(x), \beta(x) P_{Y \mid X}(\cdot \mid x)\}, \ x \not\in\{x_1^*,\dots,x_{k_x}^*\},
\end{aligned}
\end{equation*}
independently on $\tilde{p}_X$ and among themselves, where 
$$(W_{0r},W_{1r},\dots,W_{k_rr}) \sim \textsc{dirichlet}( \beta_r + k_r\sigma_r, n_{1r} - \sigma_r,\dots, n_{k_r r} - \sigma_r)$$ 
and for any $r =1,\dots,k_x$
$$\tilde{p}_{Y \mid X}^*(\cdot \mid x_r^*) \sim \textsc{py}\{\sigma_r, (\beta_r + k_r\sigma_r)P_{Y \mid X}(\cdot \mid x_r^*)\}.$$
\end{theorem}


Note that if $\sigma_r = - \beta_r/ H_r$ is strictly negative in the above theorem, then the random probability measure $\tilde{p}_{Y \mid X}^*(\cdot \mid x_r^*)$ follows a Dirichlet multinomial process with $H_r - k_r$ components, that is
\begin{equation*}
\tilde{p}_{Y \mid X}^*(\cdot \mid x_r^*) \overset{\textup{d}}{=} \sum_{j=k_r+1}^{H_r}W_{jr}^*\:\delta_{\theta_{j r}}(\cdot), 
\end{equation*}
where $(W_{k_r+1\:r}^*,\dots,W_{Hr}^*) \sim \textsc{dirichlet}(\beta_r/H_r,\dots,\beta_r/H_r)$ and $\theta_{jr} \overset{\textup{iid}}{\sim} P_{Y \mid X}(\cdot \mid x_r^*)$ for any $j=k_r + 1,\dots,H$. Hence, the posterior law of $\tilde{p}_{Y \mid X}$ is finite-dimensional, meaning that it is characterized by a finite number of random variables. On the other hand, if we set $\sigma(x) = 0$ for all $x \in \mathds{X}$, we obtain that, if $x = x_r^*$,
\begin{equation*}
\tilde{p}_{Y \mid X}(\cdot \mid x) \mid (\bm{X}^{(n)} = \bm{x}^{(n)}, \bm{Y}^{(n)} = \bm{y}^{(n)}) \sim \textsc{dp}\left(\beta_r P_{Y \mid X}(\cdot \mid x_r^*) + \sum_{j=1}^{k_r}n_{jr}\:\delta_{y_{jr}^*}(\cdot) \right),
\end{equation*}
 which implies that $\tilde{p} \mid (\bm{X}^{(n)} = \bm{x}^{(n)}, \bm{Y}^{(n)} = \bm{y}^{(n)})$ is an \textsc{edp} with updated parameters, as established in \citet{Wade2011}. 


\section{The marginal process with a discrete baseline measure}\label{sec:methods}

Recall from equations~\eqref{marginal_y}-\eqref{marginalEPY} that an exchangeable sequence $(Y_n)_{n \ge 1}$ is directed by a \emph{marginal} \textsc{epy} process if
\begin{equation*}
\begin{aligned}
Y_n \mid \tilde{p}_Y &\overset{\textup{iid}}{\sim} \tilde{p}_Y, \qquad n \ge 1, \\
\end{aligned}
\end{equation*}
where $\tilde{p}_Y(\cdot) = \tilde{p}(\mathds{X} \times \cdot)$ and $\tilde{p} \sim \textsc{epy}(\alpha P_X,\sigma(x),  \beta(x) P_{Y \mid X})$. In this section we focus on a special case of the \textsc{epy} process, arising when the marginal baseline measure $P_X$ is discrete, that is when $X_n$ takes values on a fixed set of points. More precisely, let $P_X(\cdot) = \sum_{\ell=1}^L \alpha_\ell/\alpha \:\delta_{x_\ell}(\cdot)$, with $x_1,\dots,x_L \in \mathds{X}$ and $\alpha = \sum_{\ell = 1}^L \alpha_\ell$. For the sake of the exposition we consider $L < \infty$, although our results may be easily extended to the countable case. Thus, a random probability measure $\tilde{p}_Y$ on $\mathds{Y}$ follows a marginal \textsc{epy} process with a discrete baseline measure $P_X$ if
\begin{equation}\label{marginal_discrete}
\begin{aligned}
\tilde{p}_Y(\cdot) &\overset{\textup{d}}{=} \sum_{\ell=1}^L \Pi_\ell \: \tilde{p}_\ell(\cdot) = \sum_{\ell=1}^L \sum_{h=1}^\infty \Pi_\ell \: \pi_{\ell h} \delta_{\theta_{\ell h}}(\cdot), \\
\quad (\Pi_1,\dots,\Pi_{L}) &\sim \textsc{dirichlet}(\alpha_1,\dots,\alpha_L), \qquad \tilde{p}_\ell \overset{\textup{ind}}{\sim} \textsc{py}(\sigma_\ell, \beta_\ell P_\ell),
\end{aligned}
\end{equation}
having set $\tilde{p}_\ell(\cdot) = \tilde{p}_{Y \mid X}(\cdot \mid x_\ell)$, $\theta_{\ell h} = \theta_h(x_\ell)$, $\pi_{\ell h} = \pi_h(x_\ell)$, $\sigma_\ell = \sigma(x_\ell)$, $\beta_\ell = \beta(x_\ell)$, and $P_\ell(\cdot) = P_{Y \mid X}(\cdot \mid x_\ell)$, for notational convenience. Compared to the general square-breaking representation, in the above equation the $\mathds{X}$-valued random locations $(\phi_h)_{h \ge 1}$ are replaced by the fixed values $x_1,\dots,x_L$. 

In the following, we discuss further theoretical properties and we clarify the link between the prior~\eqref{marginal_discrete} and other methods available in the literature. Interestingly, although seemingly unrelated, these proposals arise as special cases of the marginal \textsc{epy} process with a discrete $P_X$. This highlights the central role of the \textsc{epy} in a variety of contexts. In addition, making such a connection explicit allows us to develop extensions and to obtain novel modeling strategies and computational advances, that naturally arise in our unifying framework.

\subsection{Theoretical characterizations}\label{sec:discrete}

 Discrete baseline measures may have important and unexpected distributional consequences \citep{Camerlenghi2019,Lijoi2020}. In our case, such an assumption for $P_X$ leads to remarkable simplifications. To illustrate the effects of this choice, let us consider $\tilde{\mu}_X \sim \textsc{gap}(\alpha P_X)$ with $P_X(\cdot) = \sum_{\ell=1}^L \alpha_\ell/\alpha \: \delta_{x_\ell}(\cdot)$.  Then, it holds
\begin{equation*}\label{Gamma_discrete}
\tilde{\mu}_X(\cdot) \overset{\textup{d}}{=} \sum_{\ell=1}^L V_\ell \: \delta_{x_\ell}(\cdot),\qquad V_\ell \overset{\textup{ind}}{\sim} \textsc{ga}(\alpha_\ell, 1),\qquad \ell=1,\dots,L,
\end{equation*}
which is arguably a simpler representation than the general \citet{Ferguson1972} series discussed in Section~\ref{sec:background}. Indeed, the locations are  \emph{deterministic} and the jumps are independent.  As a consequence, the distribution of a marginal \textsc{epy} process $\tp_Y$ is in the form of~\eqref{marginal_discrete}.

 We now discuss characterization theorems for the joint \textsc{epy} process $\tp$ with a discrete~$P_X$, which may be used to study its distributional properties. Indeed, in this case, the Laplace functional characterizing a Gamma and Pitman--Yor random measure admits a simple expression, highlighting important connections with Cauchy-Stieltjes transforms. This is clarified in the following theorem. 

\begin{theorem}\label{teo1} Let $\tilde{\mu} \sim \textsc{ga-py}(\alpha P_X,\sigma(x),  \beta(x) P_{Y \mid X})$ with $P_X(\cdot) = \sum_{\ell=1}^L \alpha_\ell/ \alpha\:\delta_{x_\ell}(\cdot)$. Then,
\begin{equation*}\label{laplacef_discrete}
\mathds{E}\left\{e^{-\tilde{\mu}(f)}\right\} = \prod_{\ell=1}^L \mathds{E}\left[\left\{1 + \tilde{p}_{Y \mid X}(f\mid x_\ell) \right\}^{-\alpha_\ell}\right], 
\end{equation*}
where $\tilde{p}_{Y \mid X}(f\mid x) = \int_{\mathds{Y}}f(x,y)\tilde{p}_{Y \mid X}(\dd y\mid x)$ for any $x \in \mathds{X}$ and for any positive and measurable function $f : \mathds{X} \times \mathds{Y} \rightarrow \mathds{R}^+$ such that $\tilde{\mu}(f) < \infty$ almost surely.
\end{theorem}

The expectation appearing in the right hand side of the above Laplace functional is called generalized Cauchy-Stieltjes transform and it can be computed in closed form in some special cases. For example, the so-called Cifarelli-Regazzini identity \citep{Cifarelli1990} implies that if $\tilde{p} \sim \textsc{dp}(\alpha P)$ and $\tilde{\mu} \sim \textsc{gap}(\alpha P)$ then
\begin{equation}\label{cif_reg}
\mathds{E}\left[\left\{1 + \tilde{p}(f)\right\}^{-\alpha}\right] = \mathds{E}\left\{e^{-\tilde{\mu}(f)}\right\},
\end{equation}
for any positive and measurable function $f : \mathds{X} \times \mathds{Y}  \rightarrow \mathds{R}^+$ such that $\tilde{\mu}(f) < \infty$ almost surely. 
As an application of the identity in equation~\eqref{cif_reg}, we obtain the next Corollary.

\begin{corollary}\label{cor2} Let $\tilde{\mu} \sim \textsc{ga-py}(\alpha P_X, \sigma(x), \beta(x) P_{Y \mid X})$ and let $P_X(\cdot) = \sum_{\ell=1}^L \alpha_\ell/ \alpha \:\delta_{x_\ell}(\cdot)$. Moreover, assume that $\sigma(x_\ell) = 0$ and $\alpha_\ell = \beta(x_\ell)$ for any $\ell=1,\dots,L$. Then
\begin{equation*}
\mathds{E}\left\{e^{-\tilde{\mu}(f)}\right\} = \exp\left\{ - \alpha \int_{\mathds{X} \times \mathds{Y} }\log\left\{1 + f(x,y)\right\}  P(\dd x, \dd y) \right\}, 
\end{equation*}
for any positive and measurable function $f : \mathds{X} \times \mathds{Y}  \rightarrow \mathds{R}^+$ such that $\tilde{\mu}(f)  < \infty$ almost surely.
\end{corollary}

Hence, under the hypotheses of Corollary~\ref{cor2}, a \textsc{ga-py} random measure reduces to a Gamma process. In turn, this implies that an \textsc{edp} with discrete baseline measure $P_X(\cdot) = \sum_{\ell=1}^L \alpha_\ell / \alpha  \:\delta_{x_\ell}(\cdot)$ and whose parameters satisfy the constraint $\alpha_\ell = \beta(x_\ell)$ for any $\ell=1,\dots,L$, is distributed as a $\textsc{dp}(\alpha P)$.  A similar consideration was made by \citet{Wade2011}, who obtained this result by inspecting the predictive distributions. Instead, our proof relies on the \textsc{ga-py} process. Thus, again under the assumption of Corollary~\ref{cor2}, the marginal \textsc{epy} $\tilde{p}_Y$ of equation~\eqref{marginal_discrete} becomes a \textsc{dp}, namely 
\begin{equation}\label{example2}
\tilde{p}_Y(\cdot) =  \sum_{\ell=1}^L \Pi_\ell \: \tilde{p}_{Y \mid X}(\cdot \mid x_\ell) \sim \textsc{dp}\left(\sum_{\ell=1}^L \alpha_\ell P_{Y \mid X}(\cdot\mid x_\ell)\right).
\end{equation}
Therefore, the baseline measure $\sum_{\ell=1}^L \alpha_\ell/ \alpha \: P_{Y \mid X}(\cdot\mid x_\ell)$ is as a mixture.

The equivalent of the Cifarelli-Regazzini identity with $\tilde{p} \sim \textsc{py}(\sigma,\alpha P)$, has been obtained by \citet{Kerov2001} for positive $\sigma \in (0,1)$ and $\alpha > 0$. This leads to a second specialization of Theorem~\ref{teo1}, which is summarized in the following Corollary. 

\begin{corollary} \label{cor3} Let $\tilde{\mu} \sim \textsc{ga-py}(\alpha P_X, \sigma(x), \beta(x) P_{Y \mid X})$ and let $P_X(\cdot) = \sum_{\ell=1}^L \alpha_\ell / \alpha \:\delta_{x_\ell}(\cdot)$. Moreover, assume that $\sigma(x_\ell) > 0$ and $\alpha_\ell = \beta(x_\ell)$ for any $\ell=1,\dots,L$. Then
\begin{equation*}
\mathds{E}\left\{e^{-\tilde{\mu}(f)}\right\} = \prod_{\ell=1}^L \left[\int_{\mathds{Y} }\{1 + f(x_\ell,y)\}^{\sigma(x_\ell)} P_{Y \mid X}(\dd y\mid x_\ell) \right]^{-\beta(x_\ell)/\sigma(x_\ell)}, 
\end{equation*}
for any positive and measurable function $f : \mathds{X} \times \mathds{Y}  \rightarrow \mathds{R}^+$ such that $\tilde{\mu}(f)  < \infty$ almost surely. 
\end{corollary}

Corollary~\eqref{cor3} has its own theoretical interests, as it uniquely characterizes a specific \textsc{ga-py} random measure. In addition, it implies that the equivalent of equation~\eqref{example2} does not hold true in the \textsc{py} case. Specifically, under the hypothesis of Corollary~\ref{cor3}, the marginal \textsc{epy} process $\tilde{p}_Y(\cdot) =  \sum_{\ell=1}^L \Pi_\ell \: \tilde{p}_{Y \mid X}(\cdot \mid x_\ell)$, is in general not distributed as a \textsc{py}. This may still occur in a very special case, i.e. when the baseline measures and the discount parameters are equal, namely $P_{Y \mid X}(\cdot\mid x_1) = \cdots = P_{Y \mid X}(\cdot\mid x_L)$ and $\sigma(x_1) = \cdots = \sigma(x_L) > 0$, which is a known result; see e.g. Proposition 14.35 in \citet{Ghosal2017}. 

Few other generalizations of the Cifarelli-Regazzini identity are known. Specifically, consider the transform $\mathds{E}\left[\left\{1 + \tilde{p}(f)\right\}^{-\alpha}\right]$, with $\tilde{p} \sim \textsc{dp}(\beta P)$, for some positive $\beta \neq \alpha$. If such a transform were available, this would allow us to obtain the equivalent of Corollary~\ref{cor2} without imposing the constraint $\alpha_\ell = \beta(x_\ell)$. \citet{Lijoi2004} established analytic results for this transform, while \citet{James2005} provided a probabilistic interpretation of such a transform in terms of the Laplace functional of a Beta-Gamma process. However, these findings require more analytical efforts compared to~\eqref{cif_reg}.

\subsection{Discrete priors with atomic contaminations}

\textsc{dp} priors with a single atomic component are widely used in Bayesian nonparametrics. 
The common thread is the employment of a discrete random measure $\tilde{p}_Y$ on $\mathds{Y}$ having the form
\begin{equation}\label{spikeslab}
\tilde{p}_Y(\cdot) = \Pi\:\delta_{y_0}(\cdot) + (1 - \Pi)\:\tilde{p}_2(\cdot), \qquad \tilde{p}_2  \sim \textsc{dp}(\beta_2 P_2), 
\end{equation}
with $y_0 \in \mathds{Y}$ being a fixed atom, $P_2$ a diffuse probability measure, and with $\Pi \sim \textsc{beta}(\alpha_1,\alpha_2)$, independently on $\tilde{p}_2$. In some cases \citep[e.g.][]{Dunson2008b,Guindani2009,Sivaganesan2011, Cassese2019}, we have that $y_0 = 0$ and therefore $\tilde{p}_Y$ may be called a spike and slab \textsc{dp} prior. In contrast, in \citet{Scarpa2009} the atom $y_0$ is allowed to be random. Under the additional constraint $\alpha_2 = \beta_2$, the self-similarity property of the \textsc{dp} implies that model~\eqref{spikeslab} reduces to
\begin{equation}\label{spikeslab2}
\tilde{p}_Y \sim \textsc{dp}(\alpha_1 \delta_{y_0} + \alpha_2 P_2),
\end{equation}
which is the specification described in~\citet{MacLehose2007}, with $y_0 = 0$. Therefore, models~\eqref{spikeslab} and~\eqref{spikeslab2} are closely related and they are sometimes called ``outer'' and ``inner'' spike and slab, respectively.

As it may be already clear, model~\eqref{spikeslab} is a marginal \textsc{epy} process with discrete $P_X$, because it is in the form of equation~\eqref{marginal_discrete}. The equivalence between models~\eqref{spikeslab}-\eqref{spikeslab2} can be viewed as a consequence of Corollary~\ref{cor2}. This can be better appreciated by noticing that the point mass $\delta_{y_0}$ is also a trivial \textsc{dp}, namely $\delta_{y_0} = \tilde{p}_1 \sim \textsc{dp}(\alpha_1 P_1)$, with baseline measure $P_1 = \delta_{y_0}$.  In the \textsc{epy} nested clustering mechanism, the random variables $X_n$ should be interpreted as latent quantities that can only take $L = 2$ values, and the underlying baseline measure is $P_X = \alpha_1 / \alpha \:\delta_{x_1} + \alpha_2 / \alpha \:\delta_{x_2}$. In particular, each $X_n$ identifies whether the corresponding $Y_n$ has to be sampled from the  atomic contamination $\delta_{y_0}$ (i.e. $X_n = x_1$), or from the  nonparametric component $\tilde{p}_2$ (i.e. $X_n = x_2$).  


This link between model~\eqref{spikeslab} and \textsc{epy} processes leads to the following natural extension. Indeed, a simple generalization of ~\eqref{spikeslab} accounting for a more flexible clustering mechanism is
\begin{equation}\label{spyke_slab_py}
\tilde{p}_Y(\cdot) = \Pi\:\delta_{y_0}(\cdot) + (1 - \Pi)\:\tilde{p}_2(\cdot), \qquad \tilde{p}_2  \sim \textsc{py}(\sigma_2, \beta_2 P_2).
\end{equation}
Hence, the resulting $\tilde{p}_Y$ is still a marginal \textsc{epy} process and therefore it remains analytically tractable. Motivated by similar considerations, \citet{Canale2017} studied a \textsc{py} process $\bar{p}_y$ having a contaminated baseline measure, namely $\bar{p}_y \sim  \textsc{py}(\sigma, \alpha_1 \delta_{y_0} + \alpha_2 P_2)$. Importantly, Corollary~\ref{cor3} implies that $\bar{p}_y$ cannot be regarded as an \textsc{epy} process, and therefore the equivalence between inner and outer models, as in equations \eqref{spikeslab}-\eqref{spikeslab2}, is lost beyond the \textsc{dp} special case. Thus, the marginal \textsc{epy} process of equation~\eqref{spyke_slab_py} and the prior of \citet{Canale2017} are different in general, although they may be regarded as closely related  alternatives for the modeling of atomic contaminations. The main distinction is on the computational side: posterior results for $\bar{p}_y$ of \citet{Canale2017} can be derived, but their practical usage may be complicated due to the presence of cumbersome combinatorial quantities. Conversely, the posterior law of $\tilde{p}_Y$ can be readily obtained as a straightforward modification of Theorem~\ref{posterior}. The spike and slab specification in equation~\eqref{spyke_slab_py} has been independently proposed and carefully investigated by \citet{Denti2020}.

\subsection{Mixture of mixtures models}\label{sec:mixtmixt}


Bayesian nonparametric discrete priors are commonly employed for mixture modeling. In such a setting, the random variables $Z_1,\dots,Z_n$ are iid draws from a random density $\tilde{f}$, such that
\begin{equation}\label{mixture_mixture}
Z_i \mid \tilde{f} \overset{\textup{iid}}{\sim} \tilde{f}, \qquad \tilde{f}(z) = \int_{\mathds{Y}}\mathcal{K}(z \mid y)\tilde{p}_Y(\dd y), \qquad i=1,\dots,n,
\end{equation}
where $\mathcal{K}(z \mid y)$ is a kernel density and where $\tilde{p}_Y$ is a discrete random probability measure. For example, scale-location mixtures of Gaussian kernels are a  popular special case of~\eqref{mixture_mixture} when $Z_i$ is a vector on $\mathds{R}^p$. If $\tilde{p}_Y$ follows the marginal \textsc{epy} process in equation~\eqref{marginal_discrete}, then model~\eqref{mixture_mixture} may be termed a ``mixture of mixtures'' \citep{Malsiner2017}, because the random density $\tilde{f}$ becomes 
\begin{equation*}
\tilde{f}(z)\overset{\textup{d}}{=} \sum_{\ell=1}^L \Pi_\ell \int_{\mathds{Y}}\mathcal{K}(z \mid y)\tilde{p}_\ell(\dd y) =  \sum_{\ell=1}^L \sum_{h=1}^\infty \Pi_\ell \: \pi_{\ell h} \mathcal{K}(z \mid \theta_{\ell h}),
\end{equation*}
which is, indeed, a mixture model whose kernel $\int_{\mathds{Y}}\mathcal{K}(z \mid y)\tilde{p}_\ell(\dd y)$ is itself a mixture. Consistently with the nested partition mechanism described in Section~\ref{sec:urns}, there will be two levels of clustering, regulated by the latent variables $X_1,\dots,X_n$ and $Y_1,\dots,Y_n$ associated to the \textsc{epy} process. The variables $X_1,\dots,X_n$ control the global clustering, that is, they identify which kernel $\int_{\mathds{Y}}\mathcal{K}(z \mid y)\tilde{p}_\ell(\dd y)$ has to be considered. Then, conditionally on the $X_1,\dots,X_n$, the variables $Y_1,\dots,Y_n$ regulate the local clustering occurring within each kernel mixture $\int_{\mathds{Y}}\mathcal{K}(z \mid y)\tilde{p}_\ell(\dd y)$. Note that this mechanism is likely to be affected by severe identifiability issues, which may be mitigated by carefully specifying the baseline measures $P_\ell(\cdot) = P_{Y \mid X}(\cdot \mid x_\ell)$. 

Motivated by applications to the study of early-birth risk in a population of women, 
\citet{Scarpa2014} introduced model~\eqref{mixture_mixture} to cluster the $Y_1,\dots,Y_n$ subject-specific latent functions as well as predicting the $Y_{n+1}$th curve associated to a new subject. The models of~\citet{Malsiner2017} and \citet{Rigon2019} are also special cases of~\eqref{mixture_mixture}, when the discount function of the underlying \textsc{epy} process $\sigma(x)$ is strictly negative for all $x \in \mathds{X}$. 

\subsection{Dependent enriched processes}

If the random variables $Y_i^{(j)}$ are structured into groups, for unit $i = 1,\dots,n^{(j)}$ and group $j=1,\dots,d$, then the conditional iid assumption of model~\eqref{marginal_y} may be inappropriate. In fact, one may assume exchangeability only within the same group, that is
\begin{align*}
Y_n^{(j)} \mid \tilde{p}_{j,Y} & \overset{\text{iid}}{\sim} \tilde{p}_{j,Y}, \qquad n \ge 1,
\end{align*}
for $j=1,\dots,d$, with the vector of random probability measure $(\tilde{p}_{1,Y},\dots,\tilde{p}_{d,Y})$ following some prior law $\mathcal{Q}_d$. Clearly, if $\mathcal{Q}_d$ models independence across the $\tilde{p}_{j,Y}$, then the $(Y_n^{(j)})_{n \ge 1}$ would reduce to independent exchangeable sequences. 
However, this may cause an important loss of efficiency, failing in borrowing strength across groups.  Hence, it is important to specify a prior law $\mathcal{Q}_d$ that induces dependence across the random probability measures $\tilde{p}_{j,Y}$. Among the available proposals, we focus on a special case of \citet{Lijoi2014b}. Define
\begin{equation}\label{gm-Gamma}
\begin{aligned}
\tilde{p}_{j,Y}(\cdot) &=  \omega_j \tilde{q}_j(\cdot) + (1 - \omega_j)\tilde{q}_0(\cdot),  \qquad &&\omega_j \overset{\textup{iid}}{\sim} \textsc{beta}(\alpha_1, \alpha_2), \\
\tilde{q}_0 &\sim \textsc{dp}(\alpha_2 P_Y), \qquad &&\tilde{q}_j \overset{\textup{iid}}{\sim} \textsc{dp}(\alpha_1 P_Y), 
\end{aligned}
\end{equation}
independently for $j=1,\dots,d$, for some $\alpha_1, \alpha_2 > 0$, and baseline measure $P_Y$. This model is closely related to the approach of \citet{Muller2004}, who assumed a mixture distribution for $\omega_j$ with point masses at $0$ and $1$, and, in addition, that $\omega_1 = \cdots = \omega_d$. Specification~\eqref{gm-Gamma} induces dependence across groups through the presence of a common random probability measure $\tilde{q}_0$. Moreover, the self-similarity property of the \textsc{dp} implies that the marginals $\tilde{p}_{j,Y}$ are themselves \textsc{dp}s, namely $\tilde{p}_{j,Y} \sim \textsc{dp}\{(\alpha_1 + \alpha_2) P_Y\}$.

In fact, model~\eqref{gm-Gamma} is of the form~\eqref{marginal_discrete}, thus implying that each $\tilde{p}_{j,Y}$ is distributed as a marginal \textsc{edp}. Hence, extensions leveraging general \textsc{epy} processes can be envisioned. In first place, we could set 
\begin{equation*}
\tilde{q}_0 \sim \textsc{dp}(\beta_2 P_Y), \qquad \tilde{q}_j \overset{\textup{iid}}{\sim} \textsc{dp}(\beta_1 P_Y), \qquad j=1,\dots,d,
\end{equation*}
with $\alpha_1 \neq \beta_1$ and $\alpha_2 \neq \beta_2$, therefore allowing for a richer parametrization. This simple modification has quite useful implications. Indeed, it allows two random probability measures $\tilde{p}_{j,Y}$ to be highly correlated (i.e. $\alpha_2 \rightarrow \infty$) while having few number of clusters (i.e. $\beta_2 \approx 0$). This is not possible in the framework of \citet{Lijoi2014b}, where high dependence among the $\tilde{p}_{j,Y}$ is necessarily associated with a larger number of clusters. Even more generally, we could let 
\begin{equation*}
\tilde{q}_0 \sim \textsc{py}(\sigma_2, \beta_2 P_Y), \qquad \tilde{q}_j \overset{\textup{iid}}{\sim} \textsc{py}(\sigma_1, \beta_1 P_Y), \qquad j=1,\dots,d. 
\end{equation*}
This still implies that $\tilde{p}_{j,Y}$ are marginal \textsc{epy}s, and it allows the number of clusters to have different growth rates, as discussed in Section~\ref{sec:urns}. 

\section{Applications}\label{sec:applications}

We consider here two different case studies involving the \textsc{epy} process, devoting special emphasis to the elicitation or estimation of the hyperparameters $\alpha, \sigma(x), \beta(x)$ and $P$.  Indeed, as for any discrete nonparametric prior, the effect of the hyperparameters depends on the role of the random probability measure $\tilde{p}$ within the statistical model. For example, in the first case study $\tilde{p}$ is the conditional law of the \emph{observable} random vectors $(X_n, Y_n)$ and interest is in predicting the number of species within a future sample. In such a setting, the choice of the baseline measure $P$ is irrelevant as it only plays the role of labeling the different species.  In addition, we assume the diffuseness of both $P_{Y \mid X}$ and $P_X$ and we make direct usage of the predictive rule obtained in Section~\ref{sec:urns}. Instead, in the second case study we consider a mixture of mixtures model described in Section~\ref{sec:mixtmixt}, where interest is in  clustering subject-specific functions according to local  $Y_1,\dots,Y_n$ and global $X_1,\dots,X_n$ \emph{latent} features. In this case, the choices of the conditional baseline measures $P_{Y \mid X}$ and of the discrete marginal measure $P_X$ have a relevant impact on the inferential results, as it commonly occurs in mixture models. 

\subsection{The Amazonian tree flora dataset}\label{sec:app1}

The Amazonian flora is the richest assemblage of plant species on Earth. However, the exact number of tree species present in the Amazon basin, or an estimate of the proportion between common and rare species, is still unavailable. The lack of this basic information means that ecologists do not have a clear picture of the world's largest trees community.  Here we analyze the same dataset of \citet{TerSteege2013}, which is openly available online. 
Specifically, we aim at predicting the number of new species that researchers are expected to encounter in subsequent surveys, a problem with a very rich statistical literature; see e.g. \citet{Fisher1943, Good1956, Efron1976}. Refer also to \citet{Bunge1993} for a historical account on the topic. The estimation of the unobserved number of species has been more recently addressed using Bayesian nonparametrics tools, most notably in \citet{Lijoi2007b} and \citet{Favaro2009}, who spurred an interesting piece of literature focusing on species sampling models beyond the Dirichlet process. However, these approaches focus on the marginal distribution of the species, disregarding the relevant information provided, for instance, by the corresponding families. We will empirically show that enriched specifications outperform marginal models in these specific applications as they can leverage a richer parametrization and a greater amount of information.

A total of $n = 553949$ trees has been recorded in our dataset, comprising $k_y = 4962$ different species and $k_x = 115$ families of trees. The data consists in a collection of frequencies $n_{jr}$ denoting how many times the $j$th species of the $r$th family has been observed. For example, the \emph{Euterpe oleracea} species belongs to the \emph{Arecaceae} family and it has been observed $8572$ times. We let $n_r$ and  $k_r$ be the number of trees and the number of distinct species 
associated to the $r$th family, so that $\sum_{r=1}^{k_x} n_r = \sum_{r=1}^{k_x} \sum_{j=1}^{k_r} n_{jr} = n$. For instance, the \emph{Arecaceae} family comprises $70$ different species for a total of $51862$ trees. We aim at providing an accurate estimate of the number of new species $k_y(m) \in \{0, 1, \dots,m\}$ that one would observe if an additional sample of $m \ge 1$ trees were collected. The \textsc{epy} model constitutes a natural probabilistic framework for this problem. Let $X_i$ and $Y_i$ be the random variables denoting the family and the species of the $i$th tree in the sample, respectively. We assume
\begin{equation*}
\begin{aligned}
(X_i, Y_i) \mid \tilde{p} & \overset{\textup{iid}}{\sim} \tilde{p}, \qquad i \ge 1,\\
\tilde{p} &\sim \textsc{epy}(\alpha P_X,\sigma(x),  \beta(x) P_{Y \mid X}).
\end{aligned}
\end{equation*}
The distributions $P_X$ and $P_{Y \mid X}$ are \emph{diffuse} probability measures. However, they only serve as a mathematical tool for identifying ``new'' or ``old'' families and species in the urn scheme of Theorem~\ref{predictive}. Therefore, the baseline measure $P$ can be treated as a nuisance parameter. 
Note that the diffuseness of $P_X$ and $P_{Y \mid X}$ does not imply that observations $(X_i,Y_i)$ will be distinct; it only implies that new species are given new labels, as desirable. 
Still, the almost sure discreteness of $\tilde{p}$ will lead to ties among the data with positive probability. 

Recall that we are interested in the posterior distribution of $k_y(m)$ given the data, namely
\begin{equation}\label{distinct_species}
k_y(m) \mid \bm{X}^{(n)} = \bm{x}^{(n)}, \bm{Y}^{(n)} = \bm{y}^{(n)}, \qquad m \ge 1,
\end{equation}
which corresponds to the number of distinct species that one would observe within a future sample $Y_{n+1},\dots, Y_{n+m}$ and that were not observed among the data $Y_1,\dots,Y_n$. Similarly, albeit this is not the focus of this case study, one could be interested in estimating $k_x(m)$, which is the number of new families that were not observed among $X_1,\dots,X_n$. The enriched P\'olya urn scheme of Theorem~\ref{predictive} provides a very efficient way to simulate independent samples from the posterior law of~\eqref{distinct_species} by first drawing samples for $(X_{n+1},Y_{n+1}),\dots,(X_{n+m},Y_{n+m})$, given the data, and then counting the distinct values among the simulated species $Y_{n+1},\dots, Y_{n+m}$ that were not previously observed. Beside being useful for simulations, Theorem~\ref{predictive} and the associated nested Chinese restaurant metaphor shed light on the data generating process. Indeed, the families $X_1,\dots,X_n$ are generated according to the \citet{Blackwell1973} P\'olya-urn scheme and then, inside each family, the species are obtained from the urn scheme of a Pitman--Yor prior. 

\begin{figure}[tb]
\centering
\includegraphics[width=12cm]{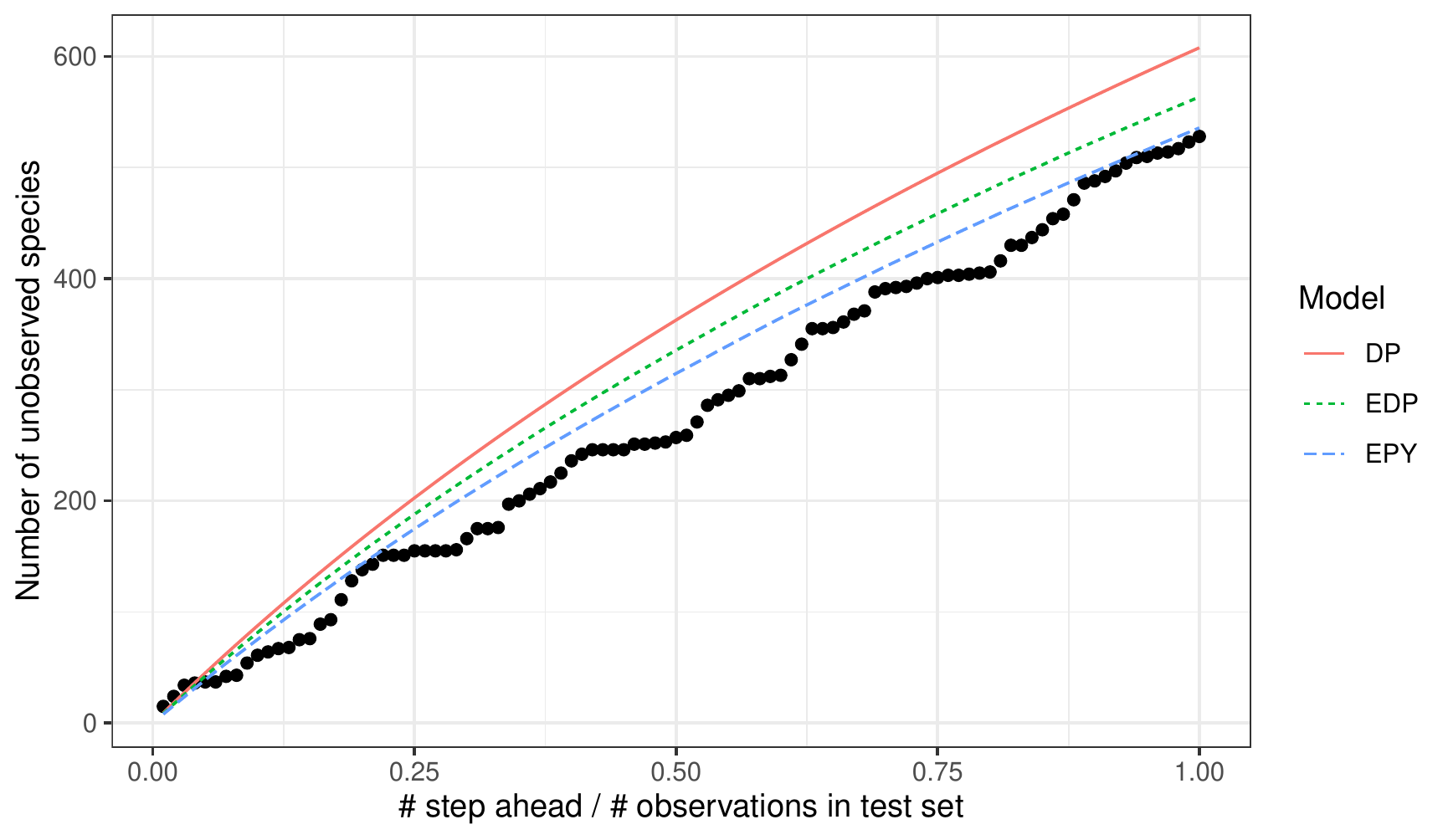}
\caption{Out-of-sample prediction for the number of new species  within the test set that were not present in the training set $\mathds{E}\{k_y(m) \mid \bm{X}^{(n_\textup{train})} = \bm{x}^{(n_\textup{train})},\bm{Y}^{(n_\textup{train})} = \bm{y}^{(n_\textup{train})}\}$, as a function of $m / n_\textup{test}$, for a \textsc{dp}, an \textsc{edp} and an \textsc{epy} model. Dots represent a sample from the test set of size $m \le n_\textup{test}$. \label{fig:model_check}}
\end{figure}

\begin{figure}[tb]
\centering
\includegraphics[width=12cm]{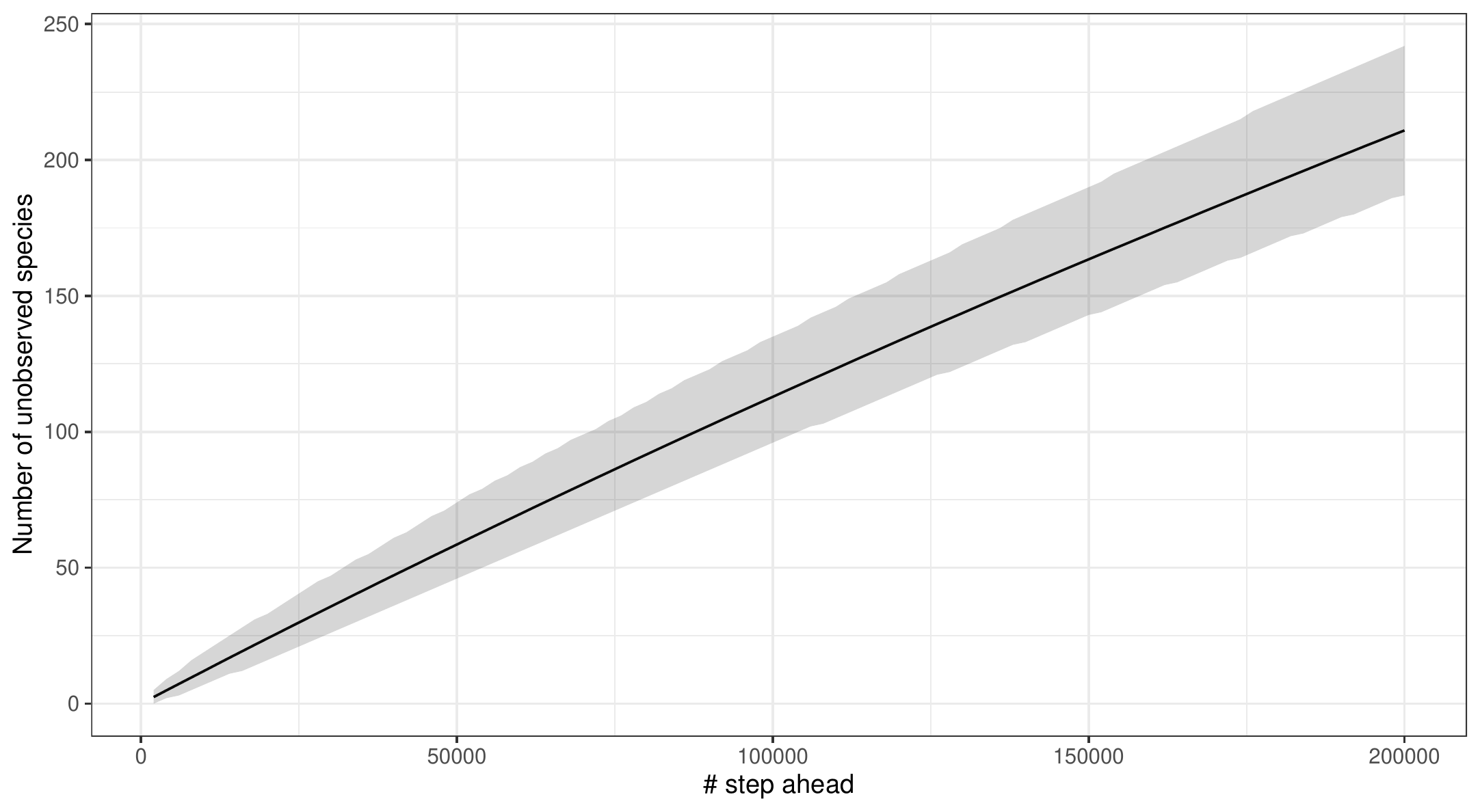}
\caption{Solid line is the prediction for the number of new species that were not present in the whole dataset $\mathds{E}\{k_y(m) \mid \bm{X}^{(n)}= \bm{x}^{(n)},\bm{Y}^{(n)}= \bm{y}^{(n)}\}$, as a function of $m$ for the \textsc{epy} model. Shaded areas represent $90\%$ pointwise credible intervals for $k_y(m) \mid \bm{X}^{(n)}= \bm{x}^{(n)},\bm{Y}^{(n)}= \bm{y}^{(n)}$. \label{fig:prediction}}
\end{figure}

To perform predictions about $k_y(m)$, we need to either specify or estimate the parameters $\alpha$, $\sigma(x)$ and $\beta(x)$. For any fixed value $x \in \mathds{X}$, the parameters $\sigma(x)$ and $\beta(x)$ are related to the growth rate and the number of species of that specific family. Since there is no natural ordering among families of trees, we let these values be themselves exchangeable. In particular, we assume that $(\sigma(x), \beta(x)) \iidsim \mathcal{Q}_{\sigma,\beta}$, that is they are independent and identically distributed  -- hence, a fortiori, exchangeable -- according to some prior law~$\mathcal{Q}_{\sigma,\beta}$. In principle, one could rely on a formal Bayesian procedure to obtain the posterior laws of $\alpha$ and each $\sigma(x)$, $\beta(x)$. In practice, this would require the implementation of a suitable Markov Chain Monte Carlo algorithm, which would heavily increase the computational burden. 
To simplify computations and following common practice in species sampling applications \citep{Lijoi2007b, Favaro2009}, we rely on a plug-in estimate for those parameters, and in particular we use their posterior mode. In other words, in what is referred to as an empirical Bayes approach in these settings, we plug-in the values $\hat{\alpha}$, $\hat{\sigma}_r = \hat{\sigma}(x^*_r)$ and $\hat{\beta}_r = \hat{\beta}(x^*_r)$ maximizing a penalized likelihood. This computationally simpler strategy might be regarded as a variational Bayes approximation of the Bayesian posterior distribution from the prior law  $\mathcal{Q}_{\sigma,\beta}$, in the lines envisaged in \citet{Zhang2020}. Proper Bayesian inference would be conceptually straightforward to implement, but the resulting Markov Chain Monte Carlo algorithm would be computationally more demanding than the proposed approximation. 

Let $\mathscr{L}(\bm{x}^{(n)}, \bm{y}^{(n)} \mid \alpha, \beta_1,\dots,\beta_{k_x},\sigma_1,\dots,\sigma_{k_x})$ be the likelihood function associated to the \textsc{epy} process, written $\mathscr{L}$ for notational simplicity. As a consequence of Theorem~\ref{predictive}, when $\sigma(x) > 0$ one can show that
\begin{equation*}
\begin{aligned}
\mathscr{L} \propto &\left[ \frac{\alpha^{k_x}}{(\alpha)_{n}} \prod_{r=1}^{k_x}(n_r - 1)! \right] \left[ \prod_{r=1}^{k_x} \frac{\prod_{j=1}^{k_r-1} (\beta_r + j\sigma_r)}{(\beta_r + 1)_{n_r-1}} \prod_{j=1}^{k_r}(1-\sigma_r)_{n_{jr}-1} \right],
\end{aligned}
\end{equation*}
where $(a)_n = a(a+1)\cdots(a+n-1)$ and $(a)_0=1$ is the Pochammer symbol. Moreover, let $f(\beta_r,\sigma_r) = f(\sigma_r)f(\beta_r)$ be the densities, for each $(\beta_r,\sigma_r)$, that are associated to the prior law $Q_{\sigma, \beta}$. In addition, let us denote with $f(\alpha)$ the prior density of $\alpha$. Thus $\hat{\alpha}$ and the pairs $(\hat{\beta}_r,\hat{\sigma}_r)$ for $r=1,\dots,k_x$ are obtained as follows
\begin{equation}\label{maximization}
\begin{aligned}
\hat{\alpha} &= \arg \max_{\alpha} f(\alpha) \frac{\alpha^{k_x}}{(\alpha)_{n}} \prod_{r=1}^{k_x}(n_r - 1)!, \\ 
(\hat{\sigma}_r, \hat{\beta}_r) &= \arg \max_{(\sigma_r,\beta_r)} f(\sigma_r,\beta_r) \frac{\prod_{j=1}^{k_r-1} (\beta_r + j\sigma_r)}{(\beta_r + 1)_{n_r-1}} \prod_{j=1}^{k_r}(1-\sigma_r)_{n_{jr}-1},
\end{aligned}
\end{equation}
for $r = 1,\dots,k_x$. Hence, the posterior modes $\hat{\alpha}$, $\hat{\sigma}_r$ and $\hat{\beta}_r$ can be easily found via numerical maximization.  As for the prior distributions $f(\alpha), f(\beta_r)$ and $f(\sigma_r)$, we let $\alpha \sim \textsc{gamma}(2,0.01)$, $\sigma_r \sim \textsc{beta}(1,10)$ and each $\beta_r \sim \textsc{gamma}(2,1)$. These prior laws regularize the otherwise ill-behaved estimates occurring when the  functions in equation~\eqref{maximization} are unbounded. The hyperparameters are selected so that the estimates $\hat{\alpha}$, $\hat{\sigma}_r$ and $\hat{\beta}_r$ are comparable with the maximum likelihood estimates in the regular cases; see the Supplementary Material for details. These prior choices implicitly assume $\sigma(x) > 0$ for any $x \in \mathds{X}$, implying that the number of species within each family is allowed to grow indefinitely with the sample size.  Finally, if a new family $X^*_{k_x+1}$ is drawn among the sampled $X_{n+1},\dots,X_{n+m}$, we then set the corresponding parameters according to the prior information, namely $(\hat{\sigma}_{k_x+1}, \hat{\beta}_{k_x+1}) = \arg \max_{(\sigma_{k_x+1},\beta_{k_x+1})} f(\sigma_{k_x+1})f(\beta_{k_x+1}) = (0,1)$. 

We validate the predictive performance of the proposed model by comparing the \textsc{epy} with a marginal \textsc{dp} on the species $Y_i$ and with an \textsc{edp}. We also considered a \textsc{py} specification; however, the obtained empirical Bayes estimate was $\sigma \approx 0$, therefore collapsing to a \textsc{dp} model.  We randomly split the dataset in a \emph{training} set and a \emph{test} set having $n_\textup{train} = 250000$ and  $n_\textup{test} = n - n_\textup{train} = 303949$ observations, respectively. The hyperparameters of the \textsc{dp} and \textsc{edp} competing models were also estimated from the data; see the Supplementary Material for the details. Conditionally on the training set, we predict the number of unobserved species $k_y(m)$, under the \textsc{dp}, \textsc{edp} and \textsc{epy} models for various choices of $m=1,\dots,n_\textup{test}$. We compare the predictions with the actual number of new species present in the full test set, if $m = n_\textup{test}$, or in a random subsample of it, when $m < n_\textup{test}$.  As apparent from Figure~\ref{fig:model_check}, all the competing methods provide a reasonable estimate for the number of unobserved species. However, the additional flexibility and information (i.e. the families) available to the enriched processes leads to more accurate predictions compared to the \textsc{dp}, with the \textsc{epy} being a slight improvement also with respect to the \textsc{edp}. 

Finally, in Figure~\ref{fig:prediction} we provide the predicted number of new species $k_y(m) \mid \bm{X}^{(n)}= \bm{x}^{(n)},\bm{Y}^{(n)}= \bm{y}^{(n)}$ within $m=1,\dots, 200000$ subsequent samples, by employing an \textsc{epy} process and conditioning on the full dataset. The expected value and  $90\%$ pointwise credible intervals are provided.  We should remark that in principle one could evaluate $k_y(m)$ for even larger values of $m$ in the attempt of estimating the total number of species present in Amazonia. However, this might lead to misleading inferential conclusions, for example because it heavily relies on the homogeneity assumptions implicit in an exchangeable model. 

\subsection{E-commerce market segmentation}\label{sec:app2}

\begin{figure}[tb]
\centering
\includegraphics[width=\textwidth]{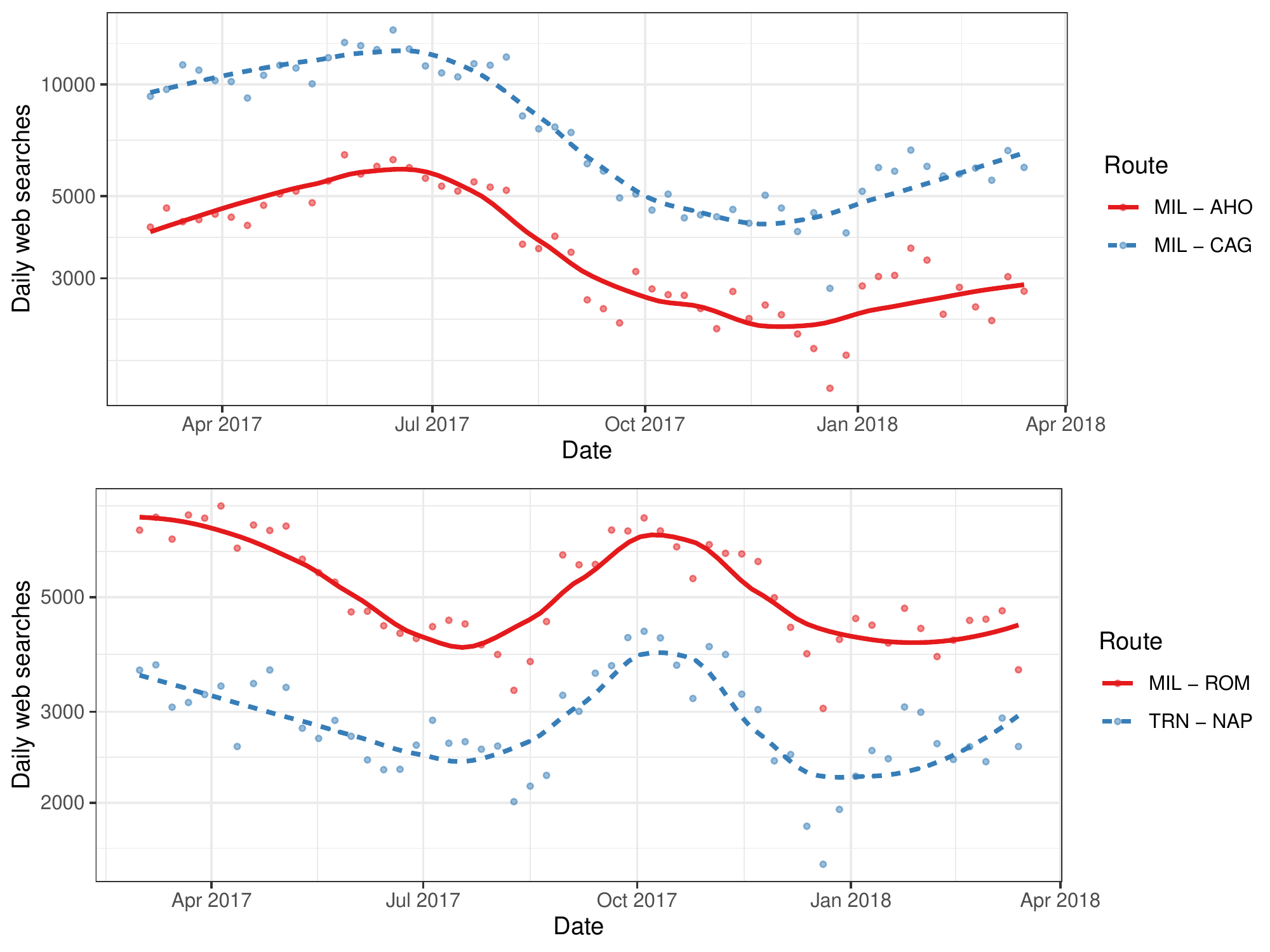}
\caption{Number of the web searches on an Italian website in the period between March 2017 and March 2018. The origin and the destination of each route are coded as follows: \texttt{MIL} = Milan, \texttt{AHO} =  Alghero, \texttt{CAG} = Cagliari, \texttt{ROM} = Rome, \texttt{TRN} = Turin, \texttt{NAP} = Naples. Smoothed trajectories are obtained using a \texttt{loess} estimate. \label{fig:flights}}
\end{figure}

An e-commerce company is interested in understanding the preferences of its customers, to implement targeted marketing strategies. This company operates online and sells flight tickets. Our analysis is based on a dataset that contains the number of times a route has been searched on the company's website, comprising a collection of weekly counts for each flight route. A few examples of these functional observations are depicted in Figure~\ref{fig:flights}. The main goal of the analyses is finding a limited number of route groups to target cluster-specific policies. Several Bayesian nonparametric approaches involving functional clustering have been recently proposed, including the works of \citet{Ray2006, Dunson2008, Bigelow2009, Petrone2009, Scarpa2014}. We here revisit the work of \citet{Rigon2019} under the lenses of general \textsc{epy} processes, which allows us
to clarify several aspects related to the elicitation of its hyperparameters. 

Let us regard the route-specific measurements $Z_i(t)$, for each route $i=1,\dots,n$ and time value $t \in \mathds{R}^+$, as error-prone realizations of unknown functions~$Y_i(t)$, that is
\begin{equation*}
\label{additive}
Z_i(t)  = Y_i(t) + \epsilon_i(t), \qquad i=1,\dots,n,
\end{equation*}
with $\epsilon_i(t)$ denoting a Gaussian random noise term, independent over flight routes and time. 
We will write $Z_i = \{Z_i(t)\}_{t \in \mathds{R}^+}$ and $Y_i = \{Y_i(t)\}_{t \in \mathds{R}^+}$ for $i=1,\dots,n$ to denote each functional observation and latent trajectory, respectively. To perform Bayesian clustering of the $Z_1,\dots,Z_n$, we let the latent trajectories $Y_1,\dots,Y_n$ to be conditionally independent draws from the marginal \textsc{epy} process described in equation~\eqref{marginal_discrete}, namely
\begin{equation*}\label{model}
Y_i \mid \tilde{p}_Y \overset{\text{iid}}{\sim} \tilde{p}_Y, \qquad i=1,\dots,n,
\end{equation*}
where $\tilde{p}_Y(\cdot) = \tilde{p}(\mathds{X} \times \cdot)$, $\tilde{p} \sim \textsc{epy}(\alpha P_X,\sigma(x),  \beta(x) P_{Y \mid X})$. Moreover, recall that we are assuming a discrete baseline measure $P_X$. This is an instance of a mixture of mixtures model, described in equation~\eqref{mixture_mixture}, for the data $Z_1,\dots,Z_n$, represented in a hierarchical manner. Note that the atoms $\theta_{\ell h} = \theta_h(x_\ell) = \{\theta_{\ell h}(t)\}_{t \in \mathds{R}^+}$ of $\tilde{p}_Y$ are themselves functions and can be interpreted as the latent trajectories of each cluster. In this specific application, a priori information is available about the shapes of these curves. For example, some flight routes are characterized by a strong cyclical component, e.g. the ones depicted in Figure~\ref{fig:flights}. The \textsc{epy} allows to naturally incorporate this prior knowledge into the model by assuming the existence of $x_1,\dots,x_L$ latent feature classes, each describing a specific functional shape (i.e. monotone, cyclical, etc.). The allocation of each function $Y_i$ to the corresponding class is regulated by the random variables $X_1,\dots,X_n$. For example, if $X_i = x_\ell$ then the $i$th latent trajectory $Y_i$ belongs to the $\ell$th feature class characterizing, say, cyclical functions. 

Compared to the first case study, the elicitation of the hyperparameters $\alpha, \sigma(x), \beta(x)$ and of the baseline measures $P_X, P_{Y \mid X}$ is more delicate.  In first place, note that each conditional base measure $P_\ell(\cdot) = P_{Y\mid X}(\cdot \mid x_\ell)$ can be regarded as a ``functional prior guess'', since for any $t \in \mathds{R}^+$ we have that $\mathds{E}\{Y_i\} =   \sum_{\ell=1}^L \alpha_\ell / \alpha \mathds{E}\{\theta_{\ell 1}\}$, with $\theta_{\ell 1} \sim P_\ell$,  for $\ell = 1,\dots,L$. Note that if inference about the latent classes $X_1,\dots,X_n$ is of interest, the measures $P_\ell$ must be ``distinguishable'', i.e. they should characterize quite different functional shapes. Otherwise, it might be difficult to infer the global clusters from the data due to severe identifiability issues. In practice, the elicitation of the conditional base measures for functional data may be based on available prior information \citep{Scarpa2014}. Consistent with the above discussion, we assume that each $\theta_{\ell h}(t)$ is linear in the parameters, with a Gaussian prior on the regression coefficients. Moreover, we let $L = 2$ and we select the conditional base measures $P_1$ and $P_2$ so that they have interpretable shapes. The first functional class ($\ell=1$) captures yearly cyclical patterns and characterizes the routes having a single peak of web-searches per year. This is the case for example of the Milan--Alghero and Milan--Cagliari routes, as apparent from Figure~\ref{fig:flights}. Hence, we let
\begin{equation*}\label{P1}
\theta_{1h}(t) = \sum_{m=1}^4 \gamma_{m1h}\mathcal{S}_m(t) +  \gamma_{51h}\cos{\left(2\pi \frac{7}{365} t\right)} + \gamma_{61h}\sin{\left(2\pi \frac{7}{365} t\right)}, 
\end{equation*}
where $\mathcal{S}_1(t),\dots,\mathcal{S}_4(t)$ are deterministic cubic spline basis functions, whose introduction allows moderate deviations from the sinusoidal component. Instead, the second functional class ($\ell=2$) characterizes functions having two peaks per year, namely
\begin{equation*}\label{P2}
\theta_{2h}(t) = \sum_{m=1}^4 \gamma_{m2h}\mathcal{S}_m(t) +\gamma_{52h}\cos{\left(2\pi \frac{14}{365} t\right)} + \gamma_{62h}\sin{\left(2\pi \frac{14}{365} t\right)}. 
\end{equation*}
The Milan--Rome and Turin--Naples routes, also depicted in Figure~\ref{fig:flights}, are potential members of this functional class. As for the prior distributions of the regression coefficients, we let $\gamma_{m\ell} \overset{\text{iid}}{\sim} \text{N}(0,1)$,  for $m=1,\dots,6$ and $\ell=1,2$, therefore inducing a quite uninformative prior, recalling that the data are standardized. 

\begin{figure}[tb]
\centering
\includegraphics[width=\textwidth]{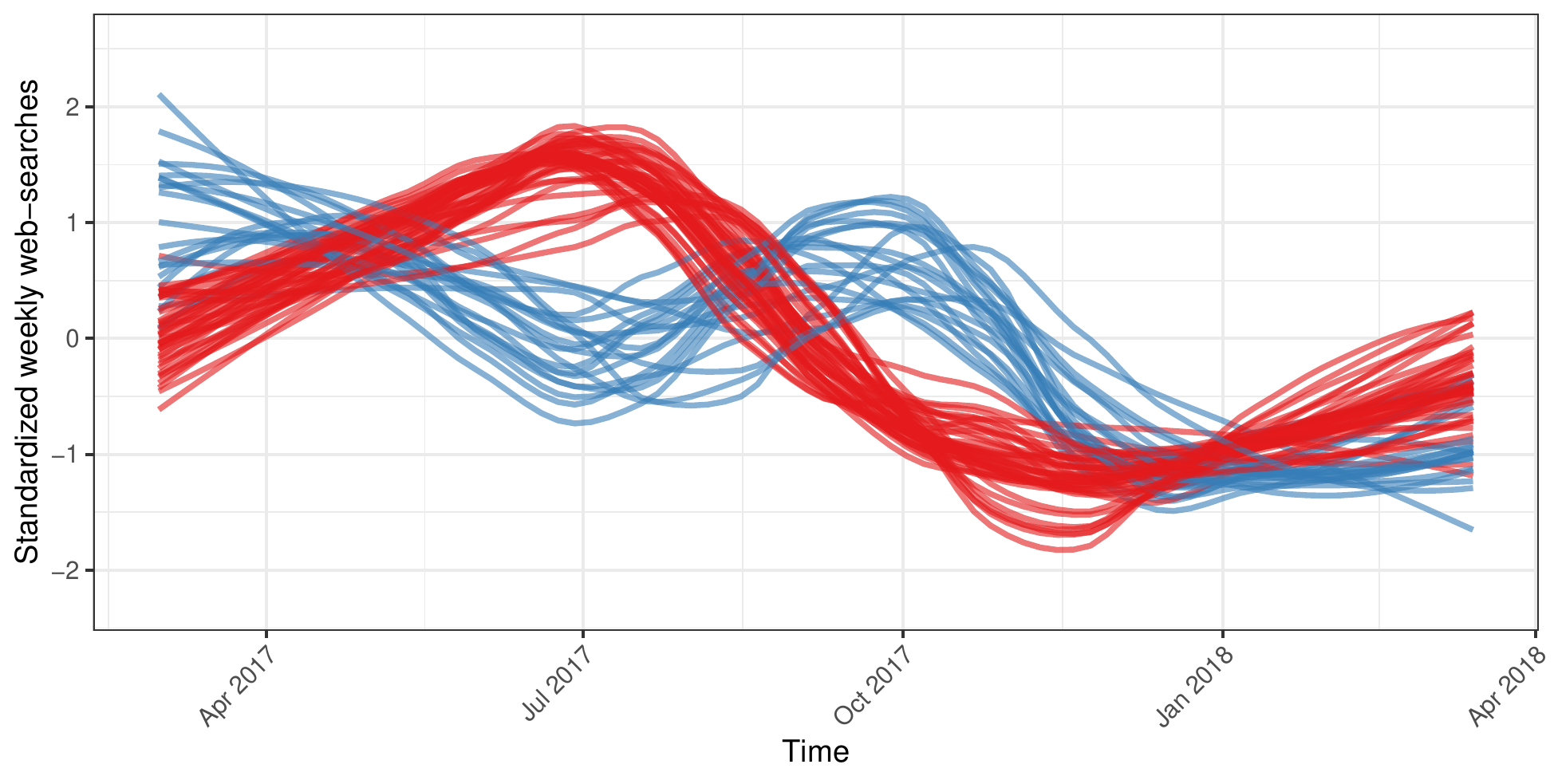}
\caption{Smoothed trajectories, obtained using a \texttt{loess} estimate, for routes in two representative clusters belonging to the first ($\ell = 1$, red lines) class and second ($\ell = 2$, blue lines) class of functions. \label{fig:clusters}}
\end{figure}

We then need to select the functions $\beta(x)$ and $\sigma(x)$. In this specific application, this amounts to the specification of four parameters $\beta_1 = \beta(x_1), \beta_2 = \beta(x_2)$ and $\sigma_1 = \sigma(x_1), \sigma_2 = \sigma(x_2)$, because the baseline measure $P_X$ is discrete and $L = 2$. Moreover, in certain clustering applications it is convenient to bound the number of groups to facilitate the practical implementation of marketing campaigns. This leads to the choice of negative values for the discount parameters $\sigma_1 = -\beta_1 / H_1$ and $\sigma_2 = -\beta_2 / H_2$, having set $H_1 = 20$ and $H_2 = 5$. In other terms, we allow no more than $20$ and $5$ clusters within each class, respectively. Moreover, we let $\beta_1 = \beta_2 = 1$ which induces a strong shrinkage on the within-class number of clusters \citep{Rousseau2011}, effectively eliminating redundant groups. Hence, the hyperparameters $\beta(x)$ and $\sigma(x)$ are selected on the basis of prior information and practical considerations. This is in contrast with the first case study of Section~\ref{sec:app1}, in which the hyperparameters $\alpha$, $\beta(x)$ and $\sigma(x)$ were estimated from the data. 

Finally, let us consider the marginal base measure $P_X$. In mixture of mixture models, the fixed values $x_1,\dots, x_L$ are arbitrary distinct labels, say the set of numbers $\{1,\dots,L\}$, each with prior proportion $\alpha_\ell/\alpha = \mathds{E}(\Pi_\ell) =  P_X(x_\ell)$. Moreover, the total mass parameter $\alpha = \sum_{\ell = 1}^L \alpha_\ell$ in this setting can be though as the prior confidence about the latter proportions. We specify a uniform prior for the functional class probabilities $(\Pi_1,\Pi_2)$, that is, we let $\alpha_1 = \alpha_2 = 1$, implying that $P_X(x_1) = P_X(x_2) = 1/2$. 

Posterior computations are performed using variational Bayes, although Markov Chain Monte Carlo methods may be considered. This results in a point estimate for the nested partition, whose outcomes are briefly summarized in the following. Two representative clusters are shown in Figure~\ref{fig:clusters}, one for each functional class $P_1$ (red lines) and $P_2$ (blue lines). As expected, the functions depicted in red are characterized by single peaked functions, while blue lines display two-peaked functions. The routes depicted in red have a peak of web searches in July, implying that the marketing manager could consider special discounts and offers in that period to incentivize sales. Similarly, marketing actions could be taken in April and subsequently in October for the group of blue routes. Clearly, in Figure~\ref{fig:clusters} we offer a description of just two specific groups as an exemplification. However, we shall remark that a total of $19$ latent trajectories have been found by the aforementioned variational Bayes algorithm, which indeed would require specific analyses and targeted marketing policies. This is carefully detailed in \citet{Rigon2019}, to which we refer for a more structured and detailed explanation in terms of migratory patterns between different geographical locations and further computational details.

\section{Concluding remarks}\label{discussion}

In this paper we extended the notion of enriched nonparametric discrete priors set forth by \citet{Wade2011}, by considering Pitman--Yor random probability measures in place of Dirichlet processes. The proposed enriched Pitman-Yor process allows for more flexibility and for nested partition structures that can include power-law behaviors; moreover, it provides a unifying probabilistic framework for several existing models within the lively Bayesian nonparametric literature. We illustrated the broad applicability of the \textsc{epy} through two different case studies, both highlighting the usefulness of nested partitions as well as the need for more flexible constructions than the \textsc{edp}. 

Our work underlines a general ``enrichment construction'' that may lead to further extensions. 
In first place, one could consider a general species sampling model for $\tp_X$, including the \textsc{py} itself, and the resulting joint random measures $\tp$ would remain well-defined. Moreover, the urn-scheme presented of Theorem~\ref{predictive} and the posterior distribution of Theorem~\ref{posterior} would be straightforward to adapt, as long as $P_X$ is diffuse. The practical advantage of using a \textsc{dp} for $\tp_X$ becomes evident whenever a \emph{discrete} marginal baseline measure $P_X$ is employed. Indeed, beyond the Dirichlet process case, the discreteness of $P_X$ entails some additional theoretical difficulties, as testified by the work of \citet{Canale2017, Camerlenghi2019}. However, if a \textsc{py} were specified for the marginal random measure $\tp_X$ in place of a \textsc{dp}, the random weights $\Pi_1,\dots,\Pi_L$ of equation~\eqref{marginal_discrete} would follow the so-called ratio-stable distribution, whose posterior law has been recently obtained by \citet{Lijoi2020}. Hence, extensions to general processes for $\tp_X$ would remain tractable even in presence of discrete baseline measures. Moreover, one could also consider more general priors for the random conditional probability measures $\tp_{Y \mid X}$, such as Gibbs-type priors \citep{DeBlasi2015} and normalized random measures with independent increments \citep{ReLP2003}. As before, the posterior properties derived in Section~\ref{sec:posterior} may be extended in the aforementioned cases. 

In another research direction, one could consider more complex nested partition mechanisms, aiming at defining random probability measures on product spaces such as $\mathds{X}_1 \times \cdots \times \mathds{X}_d$, that is allowing for more than two nesting levels. This idea has been implicitly used in \citet{Zito2022} for the development of a taxonomic classifier, but a rigorous theoretical study of the involved prior process is currently unavailable. Finally, we shall remark that covariate-dependent extensions of enriched processes can be certainly envisioned, allowing the random variables $X_n$ and $Y_n$ to depend on a set of predictors; refer to \citet{Quintana2022} for a recent and comprehensive overview. Indeed, the exchangeable assumption we made for example in Section~\ref{sec:app1} holds only in an approximate sense, because the species occurrence may depend on the spatial location, as pointed out by \citet{TerSteege2013}. These considerations encourage further theoretical investigations, for example in the development of space-dependent species sampling models, perhaps building on the contribution of \citet{Jo2017}.


\section*{Supplementary Material}


\subsection*{Proofs}

\subsubsection*{Existence of an enriched Pitman--Yor process}\label{sec:existence}

Let us denote by ${\cal B}_X$ and ${\cal B}_Y$ the Borel $\sigma$-fields associated to  $\mathds{X}$ and $\mathds{Y}$,  respectively. Denote by $\mathcal{Q}_X$ the marginal probability law of $\tp_X$, by $\mathcal{Q}_{Y \mid X}$ the joint probability law of the conditionals $\tp_{Y\mid X}(\cdot \mid x)$ for any $x \in \mathds{X}$, and by $\mathcal{Q}_{X, Y}$ the joint law $\mathcal{Q}_{X, Y} = \mathcal{Q}_X \times \mathcal{Q}_{Y \mid X}$. The proof follows the steps in \citet{Wade2011}.  First, we obtain that $\{\tp_{Y|X}(\cdot \mid x), x \in {\cal X}\}$ is a set of random {\em conditional} probability measures, as we have that  
\begin{itemize}
\item[(i)] each $\tp_{Y \mid X}(\cdot \mid x)$ is a probability measure on $\mathds{X}$ almost surely with respect to $\mathcal{Q}_{Y \mid X}$, for any $x \in \mathds{X}$;
\item[(ii)] for any Borel set $B \in {\cal B}_X$, as a function of $x$,  $\tp_{Y \mid X}(B \mid x)$ is ${\cal B}_X$-measurable, a.s. with respect to $\mathcal{Q}_{Y \mid X}$.
\end{itemize}
The property (i) is immediate, whereas (ii) is obtained from results in \citet{Ram2006} as follows. Let $\Delta$ be the set of probability measures on $\mathds{X}$ such that $\tp_{Y \mid X}(\cdot \mid x)$ is measurable as a function of $x$; results by \citet{Ram2006} ensure that, if the $\tp_{Y \mid X}( \cdot \mid x)$ are independent, then the product measure $\mathcal{Q}_{Y \mid X}$, obtained via Kolmogorov's existence theorem, assigns outer measure one to $\Delta$. 

Then, let $\mathscr{P}_D$ be the set of \emph{discrete} probability measures on $\mathds{X}$.
From the properties of the \textsc{dp}, $\tp_X$ is discrete a.s. with respect to $\mathcal{Q}_X$ and by independence of $\tp_X$ and $\tp_{Y \mid X}$ it follows that $\mathcal{Q}_{X,Y}(\mathscr{P}_D \times \Delta) = 1$. Then, results in  \citet{Ram2006} ensure that, on $\mathscr{P}_D  \times \Delta$, for any measurable subset $A \times B$ of the product space, the map $(\tp_X, \tp_{Y \mid X}) \rightarrow \int_A \tp_{Y\mid X}(B\mid x) \tp_X(\mathrm{d} x)$ is jointly measurable in $(\tp_X, \tp_{Y \mid X})$; therefore,  we can define a probability measure $\mathcal{Q}$, on the set of probability measures on the product space $\mathds{X} \times \mathds{Y}$, induced from $\mathcal{Q}_{X,Y}$ restricted to $\mathscr{P}_D  \times \Delta$ via the map  $(\tp_X, \tp_{Y \mid X}) \rightarrow \int_{( \cdot)} \tp_{Y \mid X}( \cdot \mid x) \tp_X(\mathrm{d}x)$. 


\subsubsection*{Proof of Proposition~\ref{stbr}}

The square-breaking representation of the \textsc{epy} follows directly from the stick-breaking representation of the \textsc{dp} and the \textsc{py} that has been recalled in Section~\ref{sec:background}. In particular, note that
\begin{equation*}
\begin{aligned}
\tilde{p}(A \times B) &= \int_A \tilde{p}_{Y \mid X}(B\mid x) \tilde{p}_X(\dd x) = \int_A \sum_{h=1}^\infty \pi_h(x)\:\delta_{\theta_h(x)}(B)\tilde{p}_X(\dd x)  \\
&= \sum_{\ell =1}^\infty \sum_{h=1}^\infty \xi_\ell\:\pi_h(\phi_\ell)\:\delta_{\phi_\ell}(A)\:\delta_{\theta_h(\phi_\ell)}(B),
\end{aligned}
\end{equation*}
for any Borel sets $A \subseteq \mathds{X}$ and $B \subseteq{Y}$. \qed

\subsubsection*{Proof of Theorem~\ref{laplacef_discrete}}
From equation~\eqref{laplace_Gamma} and recalling Definition~\ref{def1} we obtain that the Laplace functional of a Gamma and Pitman--Yor random measure $\tilde{\mu}$ can be written as
\begin{equation*}\label{laplacef}
\mathds{E}\left\{e^{-\tilde{\mu}(f)}\right\} = \mathds{E}\left[\exp\left\{-\alpha \int_{\mathds{X}}\log\{1 + \tilde{p}_{Y \mid X}(f\mid x)\}P_X(\dd x)\right\}\right],
\end{equation*}
where $\tilde{p}_{Y \mid X}(f\mid x) = \int_{\mathds{Y}}f(x,y)\tilde{p}_{Y \mid X}(\dd y\mid x)$ and for any positive and measurable function $f : \mathds{X} \times \mathds{Y} \rightarrow \mathds{R}^+$ such that $\tilde{\mu}(f) < \infty$ almost surely. The above Laplace functional is fully general and it does not require further restrictions on $P_X$. Then, by exploiting the discreteness of $P_X$ and the independence among the conditional laws $\tilde{p}_{Y \mid X}$ we obtain
\begin{equation*}
\begin{aligned}
\mathds{E}\left\{e^{-\tilde{\mu}(f)}\right\} &= \mathds{E}\left(\exp\left[- \sum_{\ell=1}^L \alpha_\ell \log\{1 + \tilde{p}_{Y \mid X}(f\mid x_\ell)\}\right]\right) \\
&= \prod_{\ell=1}^L \mathds{E}\left[\left\{1 + \tilde{p}_{Y \mid X}(f\mid x_\ell) \right\}^{-\alpha_\ell}\right],
\end{aligned}
\end{equation*}
which concludes the proof for $L < \infty$. This result can be easily extended to the countable case as an application of dominated convergence theorem. \qed

\subsubsection*{Proof of Theorem~\ref{predictive}}
By definition of the \textsc{epy} and from equation~\eqref{enriched_sampling}, we get that the marginal law $\tilde{p}_X$ is independent on $\bm{Y}^{(n)}$, given $\bm{X}^{(n)}= \bm{x}^{(n)}$, so that for any Borel set $A \subseteq \mathds{X}$
\begin{equation*}
\begin{aligned}
\mathds{P}(X_{n+1} \in A \mid \bm{X}^{(n)}&= \bm{x}^{(n)}, \bm{Y}^{(n)}= \bm{y}^{(n)}) =\\
& \qquad\qquad= \mathds{E}(\tilde{p}_X(A) \mid \bm{X}^{(n)}= \bm{x}^{(n)}, \bm{Y}^{(n)}= \bm{y}^{(n)})\\ 
& \qquad\qquad= \mathds{E}(\tilde{p}_X(A) \mid \bm{X}^{(n)}= \bm{x}^{(n)})\\
& \qquad\qquad=  \mathds{P}(X_{n+1} \in A \mid \bm{X}^{(n)}= \bm{x}^{(n)}),
\end{aligned}
\end{equation*}
which leads to the \citet{Blackwell1973} scheme for the exchangeable sequence $(X_n)_{n \ge 1}$. Note that the (potential) discreteness of $P_X$ poses no issues here. The second part of the theorem is obtained  by exploiting the independence among the \textsc{py} conditional laws $\tilde{p}_{Y \mid X}$. Indeed, each subset of observations $\bm{Y}^{(n_r)}$ follows the well-known scheme of the \textsc{py}, which is described e.g. in \citet{Pitman1996} when $P_{Y \mid X}$ is diffuse. \qed

We now show that, in fact, the predictive scheme in Theorem~\ref{predictive} provides a characterization of the \textsc{epy} process, as claimed in Remark~\ref{rem1}. By Ionescu-Tulcea theorem, the sequence of predictive distributions,  say $\mathds{P}(Z_{n+1} \in \cdot \mid Z_1=z_1, \ldots, Z_n = z_n)$, for $n \geq 1$ uniquely characterizes the probability law of the stochastic process $(Z_n)_{n \geq 1}$.  Therefore, the sequence of the predictive distributions \eqref{pred1}-\eqref{pred2} characterizes the probability law, say $\mathcal{P}$, of the stochastic process $\left( (X_n, Y_n)\right)_{n \ge 1}$. On the other hand, an exchangeable probability law with an \textsc{epy} directing measure leads to the predictive rule \eqref{pred1}-\eqref{pred2}, as shown in Theorem~\eqref{predictive}. By unicity, the law $\mathcal{P}$ necessarily coincides with such an exchangeable law.  

\subsubsection*{Proof of Theorem~\ref{posterior}}

The independence among $\tilde{p}_X$ and $\bm{Y}^{(n)}$, given $\bm{X}^{(n)}= \bm{x}^{(n)}$ immediately leads to the first part of the theorem, thank to conjugacy of the \textsc{dp} \citep{Ferguson1973}. Similarly, the posterior distribution of each conditional law $\tilde{p}_{Y \mid X}$, thanks to their independence, is obtained as an application of Corollary~20 in~\citet{Pitman1996} to each subset of observations~$\bm{Y}^{(n_r)}$, which requires the diffuseness of $P_{Y \mid X}$. \qed

\subsection*{Choice of the hyperparameters}

We discuss here in more detail the prior choices for the three competing models considered in Section~\ref{sec:app1}. We compare the estimates $\hat{\alpha}$, $\hat{\sigma}_r$ and $\hat{\beta}_r$ that results from these prior choices with the corresponding maximum likelihood estimates. We show that in most cases the penalization has little effect and therefore it mainly serves as a tool to regularize ill-behaved estimates.

\begin{table}[ptb]
\centering
\begin{tabular}{rrrrrrrrrrr}
  \toprule
$r$ & 1 & 2 & 3 & 4 & 5 & 6 & 7 & 8 & 9 & 10 \\ 
  \midrule
  $\hat{\beta}_r$ & 0.53 & 2.11 & 0.24 & 1.00 & 4.37 & 0.36 & 41.19 & 17.29 & 2.41 & 2.42 \\ 
  $\hat{\beta}_{r,\textsc{mle}}$ & 0.34 & 2.44 & 0.00 & 1.00 & 5.03 & 0.21 & 50.66 & 20.94 & 2.81 & 2.75 \\ 
\bottomrule
\end{tabular}
\caption{\textsc{edp} model. Penalized estimates $\hat{\beta}_r$ and maximum likelihood estimates $\hat{\beta}_{r,\textsc{mle}}$ for a subset of coefficients, with $r=1,\dots,10$. \label{tab1bis}}
\end{table}

\begin{table}[ptb]
\centering
\begin{tabular}{rrrrrrr}
\toprule
 & Minimum & 1st Quartile & Median & Mean & 3rd Quartile & Maximum \\ 
\midrule
$\hat{\beta}_r$ & 0.12 & 0.63 & 1.55 & 6.19 & 5.73 & 104.39 \\ 
\midrule
$\hat{\beta}_{r,\textsc{mle}}$ & 0.00 & 0.53 & 2.01 & 7.41 & 6.89 & 125.84 \\ 
\bottomrule
\end{tabular}
\caption{\textsc{edp} model. Summary of the penalized estimates $\hat{\beta}_r$ and of the maximum likelihood estimates $\hat{\beta}_{r,\textsc{mle}}$. \label{tab1}}
\end{table}

\begin{table}[ptb]
\centering
\begin{tabular}{rrrrrrr}
\toprule
 & Minimum & 1st Quartile & Median & Mean & 3rd Quartile & Maximum \\ 
\midrule
$\hat{\beta}_r$ & 0.31 & 0.87 & 2.06 & 4.56 & 4.68 & 56.80 \\ 
$\hat{\sigma}_r$ & 0.00 & 0.00 & 0.00 & 0.03 & 0.02 & 0.27 \\ 
\bottomrule
\end{tabular}
\caption{\textsc{epy} model. Summary of the penalized estimates $\hat{\beta}_r$ and $\hat{\sigma}_r$.\label{tab2}}
\end{table}

\subsubsection*{DP model}

In the \textsc{dp} model we need to specify the so-called precision parameter, say $\beta$, and we let $\beta \sim \textsc{gamma}(2,0.001)$. The consequent penalized likelihood approach is $\hat{\beta} = 765.53$, whereas the maximum likelihood estimate is $\hat{\beta}_\textsc{mle} = 765.48$. Hence, the prior penalty has almost no effect in this specific case.

\subsubsection*{EDP and EPY models}

The hyperparameter settings for the the \textsc{edp} and the \textsc{epy} models are similar. For the \textsc{edp}, we use $\alpha \sim \textsc{gamma}(2,0.01)$, $\sigma_r = 0$, and $\beta_r \sim \textsc{gamma}(2,1)$. For the \textsc{epy}, we let $\alpha \sim \textsc{gamma}(2,0.01)$, $\sigma_r \sim \textsc{beta}(1,10)$, and $\beta_r \sim \textsc{gamma}(2,1)$. Note that the estimates for $\alpha$ coincide in these two models and one gets $\hat{\alpha} = 11.34$ and $\hat{\alpha}_\textsc{mle} = 11.24$.  Hence, the prior penalty has almost no effect in the estimation of $\alpha$.

Viceversa, we expect the prior to have an effect in the estimation of $\beta_r$ and $\sigma_r$. As an illustration, we report in Table~\ref{tab1bis} the penalized estimates $\hat{\beta}_r$, for $r=1,\dots,10$,  for the \textsc{edp} model and we compare them with the corresponding maximum likelihood estimates $\hat{\beta}_{r,\textsc{mle}}$. A summary of all the estimates for $r=1,\dots,k_x$ is given in Table~\ref{tab1}. Note that the expected value of the $\textsc{gamma}(2,1)$ prior roughly corresponds to the median of the maximum likelihood estimates. As expected, the prior on $\beta_r$ shrinks the maximum likelihood estimates towards $2$ and avoids ill-behaved maximization problems. For example, without the penalty term one would get $\hat{\beta}_{3,\textsc{mle}} = 0$, which is not an admissible value. 

Finally, we report the summary of the estimates $(\hat{\beta}_r,\hat{\sigma}_r)$ for the \textsc{epy} model in Table~\ref{tab2}. The prior on $\sigma_r$ shrinks the estimates $\hat{\sigma}_r$ towards $0$ and therefore the corresponding $\beta_r$ parameters are quite similar to those of the \textsc{edp}.

\bibliographystyle{chicago}
\bibliography{biblio}

\begin{thebibliography}{}

\bibitem[\protect\citeauthoryear{Bigelow and Dunson}{Bigelow and
  Dunson}{2009}]{Bigelow2009}
Bigelow, J.~L. and D.~B. Dunson (2009).
\newblock {Bayesian semiparametric joint models for functional predictors}.
\newblock {\em Journal of the American Statistical Association\/}~{\em
  104\/}(485), 26--36.

\bibitem[\protect\citeauthoryear{Blackwell and MacQueen}{Blackwell and
  MacQueen}{1973}]{Blackwell1973}
Blackwell, D. and J.~B. MacQueen (1973).
\newblock {Ferguson distributions via P\'olya urn schemes}.
\newblock {\em The Annals of Statistics\/}~{\em 1\/}(2), 353--355.

\bibitem[\protect\citeauthoryear{Bunge and Fitzpatrick}{Bunge and
  Fitzpatrick}{1993}]{Bunge1993}
Bunge, J. and M.~Fitzpatrick (1993).
\newblock Estimating the number of species: a review.
\newblock {\em Journal of the American Statistical Association\/}~{\em
  88\/}(421), 364--73.

\bibitem[\protect\citeauthoryear{Camerlenghi, Lijoi, Orbanz, and
  Pr{\"{u}}nster}{Camerlenghi et~al.}{2019}]{Camerlenghi2019}
Camerlenghi, F., A.~Lijoi, P.~Orbanz, and I.~Pr{\"{u}}nster (2019).
\newblock {Distribution theory for hierarchical processes}.
\newblock {\em The Annals of Statistics\/}~{\em 47\/}(1), 67--92.

\bibitem[\protect\citeauthoryear{Canale, Lijoi, Nipoti, and
  Pr{\"{u}}nster}{Canale et~al.}{2017}]{Canale2017}
Canale, A., A.~Lijoi, B.~Nipoti, and I.~Pr{\"{u}}nster (2017).
\newblock {On the Pitman-Yor process with spike and slab base measure}.
\newblock {\em Biometrika\/}~{\em 104\/}(3), 681--697.

\bibitem[\protect\citeauthoryear{Cassese, Zhu, Guindani, and Vannucci}{Cassese
  et~al.}{2019}]{Cassese2019}
Cassese, A., W.~Zhu, M.~Guindani, and M.~Vannucci (2019).
\newblock {A Bayesian nonparametric spiked process prior}.
\newblock {\em Bayesian Analysis\/}~{\em 14\/}(2), 553--572.

\bibitem[\protect\citeauthoryear{Cifarelli and Regazzini}{Cifarelli and
  Regazzini}{1990}]{Cifarelli1990}
Cifarelli, D.~M. and E.~Regazzini (1990).
\newblock {Distribution functions of means of a Dirichlet process}.
\newblock {\em The Annals of Statistics\/}~{\em 18\/}(1), 429--442.

\bibitem[\protect\citeauthoryear{Connor and Mosimman}{Connor and
  Mosimman}{1969}]{Connor1969}
Connor, R.~J. and J.~E. Mosimman (1969).
\newblock {Concepts of independence for proportions with generalization of the
  Dirichlet distribution}.
\newblock {\em Journal of the American Statistical Association\/}~{\em
  64\/}(325), 194-- 206.

\bibitem[\protect\citeauthoryear{Consonni and Veronese}{Consonni and
  Veronese}{2001}]{Consonni2001}
Consonni, G. and P.~Veronese (2001).
\newblock Conditionally reducible natural exponential families and enriched
  conjugate priors.
\newblock {\em Scandinavian Journal of Statistics\/}~{\em 28}, 377--406.

\bibitem[\protect\citeauthoryear{{De Blasi}, Favaro, Lijoi, Mena,
  Pr{\"{u}}nster, and Ruggiero}{{De Blasi} et~al.}{2015}]{DeBlasi2015}
{De Blasi}, P., S.~Favaro, A.~Lijoi, R.~H. Mena, I.~Pr{\"{u}}nster, and
  M.~Ruggiero (2015).
\newblock {Are Gibbs-type priors the most natural generalization of the
  Dirichlet process?}
\newblock {\em IEEE Transactions on Pattern Analysis and Machine
  Intelligence\/}~{\em 37\/}(2), 212--229.

\bibitem[\protect\citeauthoryear{Denti}{Denti}{2020}]{Denti2020}
Denti, F. (2020).
\newblock {\em Bayesian mixtures for large scale inference}.
\newblock Ph.\ D. thesis, University of Milan Bicocca.

\bibitem[\protect\citeauthoryear{Doksum}{Doksum}{1974}]{Doksum1974}
Doksum, K. (1974).
\newblock {Tailfree and neutral random probabilities and their posterior
  distributions}.
\newblock {\em The Annals of Probability\/}~{\em 2\/}(2), 183--201.

\bibitem[\protect\citeauthoryear{Dunson, Herring, and Engel}{Dunson
  et~al.}{2008}]{Dunson2008b}
Dunson, D.~B., A.~H. Herring, and S.~M. Engel (2008).
\newblock {Bayesian selection and clustering of polymorphisms in functionally
  related genes}.
\newblock {\em Journal of the American Statistical Association\/}~{\em
  103\/}(482), 534--546.

\bibitem[\protect\citeauthoryear{Dunson, Herring, and Siega-Riz}{Dunson
  et~al.}{2008}]{Dunson2008}
Dunson, D.~B., A.~H. Herring, and A.~M. Siega-Riz (2008).
\newblock {Bayesian inference on changes in response densities over predictor
  clusters}.
\newblock {\em Journal of the American Statistical Association\/}~{\em
  103\/}(484), 1508--1517.

\bibitem[\protect\citeauthoryear{Efron and Thisted}{Efron and
  Thisted}{1976}]{Efron1976}
Efron, B. and R.~Thisted (1976).
\newblock {Estimating the number of unseen species: How many words did
  Shakespeare know?}
\newblock {\em Biometrika\/}~{\em 63\/}(3), 435--447.

\bibitem[\protect\citeauthoryear{Favaro, Lijoi, Mena, and
  Pr{\"{u}}nster}{Favaro et~al.}{2009}]{Favaro2009}
Favaro, S., A.~Lijoi, R.~H. Mena, and I.~Pr{\"{u}}nster (2009).
\newblock {Bayesian non-parametric inference for species variety with a
  two-parameter Poisson-Dirichlet process prior}.
\newblock {\em Journal of the Royal Statistical Society. Series B: Statistical
  Methodology\/}~{\em 71\/}(5), 993--1008.

\bibitem[\protect\citeauthoryear{Ferguson}{Ferguson}{1973}]{Ferguson1973}
Ferguson, T.~S. (1973).
\newblock {A Bayesian analysis of some nonparametric problems}.
\newblock {\em The Annals of Statistics\/}~{\em 1\/}(2), 209--230.

\bibitem[\protect\citeauthoryear{Ferguson and Klass}{Ferguson and
  Klass}{1972}]{Ferguson1972}
Ferguson, T.~S. and M.~J. Klass (1972).
\newblock {A representation of independent increment processes without Gaussian
  components}.
\newblock {\em Annals of Mathematical Statistics\/}~{\em 43\/}(5), 1634--1643.

\bibitem[\protect\citeauthoryear{Fisher, Corbet, and Williams}{Fisher
  et~al.}{1943}]{Fisher1943}
Fisher, R.~A., A.~S. Corbet, and C.~B. Williams (1943).
\newblock The relation between the number of species and the number of
  individuals in a random sample of an animal population.
\newblock {\em Journal of Animal Ecology\/}~{\em 12\/}(1), 42--58.

\bibitem[\protect\citeauthoryear{Gadd, Wade, and Boukouvalas}{Gadd
  et~al.}{2019}]{Gadd2019}
Gadd, C.~W., S.~Wade, and A.~Boukouvalas (2019).
\newblock Enriched mixtures of gaussian process experts.
\newblock {\em arXiv:1905.12969\/}, 1--10.

\bibitem[\protect\citeauthoryear{Ghosal and van~der Vaart}{Ghosal and van~der
  Vaart}{2017}]{Ghosal2017}
Ghosal, S. and A.~van~der Vaart (2017).
\newblock {\em Fundamentals of Nonparametric Bayesian Inference}.
\newblock Cambridge Series in Statistical and Probabilistic Mathematics.

\bibitem[\protect\citeauthoryear{Good and Toulmin}{Good and
  Toulmin}{1956}]{Good1956}
Good, I.~J. and G.~H. Toulmin (1956).
\newblock The number of new species, and the increase in population coverage,
  when a sample is increased.
\newblock {\em Biometrika\/}~{\em 43\/}(1-2), 45--63.

\bibitem[\protect\citeauthoryear{Guindani, M\"uller, and Zhang}{Guindani
  et~al.}{2009}]{Guindani2009}
Guindani, M., P.~M\"uller, and S.~Zhang (2009).
\newblock {A Bayesian discovery procedure}.
\newblock {\em Journal of the Royal Statistical Society. Series B: Statistical
  Methodology\/}~{\em 71\/}(5), 905--925.

\bibitem[\protect\citeauthoryear{Ishwaran and James}{Ishwaran and
  James}{2001}]{Ishwaran2001}
Ishwaran, H. and L.~F. James (2001).
\newblock {Gibbs sampling methods for stick-breaking priors}.
\newblock {\em Journal of the American Statistical Association\/}~{\em
  96\/}(453), 161--173.

\bibitem[\protect\citeauthoryear{James}{James}{2005}]{James2005}
James, L.~F. (2005).
\newblock {Functionals of Dirichlet processes, the Cifarelli--Regazzini
  identity and Beta-Gamma processes}.
\newblock {\em The Annals of Statistics\/}~{\em 33\/}(2), 647--660.

\bibitem[\protect\citeauthoryear{Jo, Lee, M\"uller, Quintana, and Trippa}{Jo
  et~al.}{2017}]{Jo2017}
Jo, S., J.~Lee, P.~M\"uller, F.~Quintana, and L.~Trippa (2017).
\newblock Dependent species sampling models for spatial density estimation.
\newblock {\em Bayesian Analysis\/}~{\em 12\/}(2), 379--406.

\bibitem[\protect\citeauthoryear{Kerov and Tsilevich}{Kerov and
  Tsilevich}{2001}]{Kerov2001}
Kerov, S.~V. and N.~V. Tsilevich (2001).
\newblock {The Markov-Krein correspondence in several dimensions}.
\newblock {\em Zapisky Nauchnykh Seminarov POMI\/}~{\em 283}, 98--122.

\bibitem[\protect\citeauthoryear{Lijoi, Mena, and Pr{\"{u}}nster}{Lijoi
  et~al.}{2007a}]{Lijoi2007b}
Lijoi, A., R.~H. Mena, and I.~Pr{\"{u}}nster (2007a).
\newblock {Bayesian nonparametric estimation of the probability of discovering
  new species}.
\newblock {\em Biometrika\/}~{\em 94\/}(4), 769--786.

\bibitem[\protect\citeauthoryear{Lijoi, Mena, and Pr{\"{u}}nster}{Lijoi
  et~al.}{2007b}]{Lijoi2007}
Lijoi, A., R.~H. Mena, and I.~Pr{\"{u}}nster (2007b).
\newblock {Controlling the reinforcement in Bayesian non-parametric mixture
  models}.
\newblock {\em Journal of the Royal Statistical Society. Series B: Statistical
  Methodology\/}~{\em 69\/}(4), 715--740.

\bibitem[\protect\citeauthoryear{Lijoi, Nipoti, and Pr{\"{u}}nster}{Lijoi
  et~al.}{2014}]{Lijoi2014b}
Lijoi, A., B.~Nipoti, and I.~Pr{\"{u}}nster (2014).
\newblock {Bayesian inference with dependent normalized completely random
  measures}.
\newblock {\em Bernoulli\/}~{\em 20\/}(3), 1260--1291.

\bibitem[\protect\citeauthoryear{Lijoi and Pr{\"{u}}nster}{Lijoi and
  Pr{\"{u}}nster}{2010}]{Lijoi2010}
Lijoi, A. and I.~Pr{\"{u}}nster (2010).
\newblock {Models beyond the Dirichlet process}.
\newblock In N.~L. Hjort, C.~C. Holmes, P.~Muller, and S.~G. Walker (Eds.),
  {\em Bayesian Nonparametrics}. Cambridge University Press.

\bibitem[\protect\citeauthoryear{Lijoi, Pr{\"{u}}nster, and Rigon}{Lijoi
  et~al.}{2020}]{Lijoi2020}
Lijoi, A., I.~Pr{\"{u}}nster, and T.~Rigon (2020).
\newblock {The Pitman--Yor multinomial model for mixture modeling}.
\newblock {\em Biometrika\/}~{\em 107\/}(4), 891--906.

\bibitem[\protect\citeauthoryear{Lijoi and Regazzini}{Lijoi and
  Regazzini}{2004}]{Lijoi2004}
Lijoi, A. and E.~Regazzini (2004).
\newblock {Means of a Dirichlet process and multiple hypergeometric functions}.
\newblock {\em The Annals of Applied Probability\/}~{\em 32\/}(2), 1469--1495.

\bibitem[\protect\citeauthoryear{MacLehose, Dunson, Herring, and
  Hopping}{MacLehose et~al.}{2007}]{MacLehose2007}
MacLehose, R.~F., D.~B. Dunson, A.~H. Herring, and J.~A. Hopping (2007).
\newblock Bayesian methods for highly correlated exposure data.
\newblock {\em Epidemiology\/}~{\em 18\/}(2), 199--207.

\bibitem[\protect\citeauthoryear{Malsiner-Walli, Fr{\"{u}}hwirth-Schnatter, and
  Gr{\"{u}}n}{Malsiner-Walli et~al.}{2017}]{Malsiner2017}
Malsiner-Walli, G., S.~Fr{\"{u}}hwirth-Schnatter, and B.~Gr{\"{u}}n (2017).
\newblock {Identifying mixtures of mixtures using Bayesian estimation}.
\newblock {\em Journal of Computational and Graphical Statistics\/}~{\em
  26\/}(2), 285--295.

\bibitem[\protect\citeauthoryear{M\"uller, Quintana, and Rosner}{M\"uller
  et~al.}{2004}]{Muller2004}
M\"uller, P., F.~Quintana, and G.~Rosner (2004).
\newblock {A method for combining inference across related nonparametric
  Bayesian models}.
\newblock {\em Journal of the Royal Statistical Society. Series B: Statistical
  Methodology\/}~{\em 66\/}(3), 735--749.

\bibitem[\protect\citeauthoryear{Perman, Pitman, and Yor}{Perman
  et~al.}{1992}]{Perman1992}
Perman, M., J.~Pitman, and M.~Yor (1992).
\newblock {Size-biased sampling of Poisson point processes and excursions}.
\newblock {\em Probability Theory and Related Fields\/}~{\em 92}, 21--39.

\bibitem[\protect\citeauthoryear{Petrone, Guindani, and Gelfand}{Petrone
  et~al.}{2009}]{Petrone2009}
Petrone, S., M.~Guindani, and A.~E. Gelfand (2009).
\newblock {Hybrid Dirichlet mixture models for functional data}.
\newblock {\em Journal of the Royal Statistical Society. Series B: Statistical
  Methodology\/}~{\em 71\/}(4), 755--782.

\bibitem[\protect\citeauthoryear{Pitman}{Pitman}{1996}]{Pitman1996}
Pitman, J. (1996).
\newblock {Some developments of the Blackwell-Macqueen urn scheme}.
\newblock {\em Statistics, Probability and Game Theory\/}~{\em 30}, 245--267.

\bibitem[\protect\citeauthoryear{Pitman and Yor}{Pitman and
  Yor}{1997}]{Pitman1997}
Pitman, J. and M.~Yor (1997).
\newblock {The two-parameter Poisson-Dirichlet distribution derived from a
  stable subordinator}.
\newblock {\em The Annals of Probability\/}~{\em 25\/}(2), 855--900.

\bibitem[\protect\citeauthoryear{Quintana, M{\"{u}}ller, Jara, and
  MacEachern}{Quintana et~al.}{2022}]{Quintana2022}
Quintana, F., P.~M{\"{u}}ller, A.~Jara, and S.~V. MacEachern (2022).
\newblock {The dependent Dirichlet process and related models}.
\newblock {\em Statistical Science\/}~{\em 37\/}(1), 24--41.

\bibitem[\protect\citeauthoryear{Ramamoorthi and Sangalli}{Ramamoorthi and
  Sangalli}{2006}]{Ram2006}
Ramamoorthi, R. and L.~Sangalli (2006).
\newblock {On a characterization of Dirichlet distribution}.
\newblock In {\em Proceedings of the International Conference on Bayesian
  Statistics and its Applications}, pp.\  385--397.

\bibitem[\protect\citeauthoryear{Ray and Mallick}{Ray and
  Mallick}{2006}]{Ray2006}
Ray, S. and B.~Mallick (2006).
\newblock {Functional clustering by Bayesian wavelet methods}.
\newblock {\em Journal of the Royal Statistical Society. Series B: Statistical
  Methodology\/}~{\em 68\/}(2), 305--332.

\bibitem[\protect\citeauthoryear{Regazzini, Lijoi, and
  Pr{\"{u}}nster}{Regazzini et~al.}{2003}]{ReLP2003}
Regazzini, E., A.~Lijoi, and I.~Pr{\"{u}}nster (2003).
\newblock {Distributional results for means of normalized random measures with
  independent increments}.
\newblock {\em The Annals of Statistics\/}~{\em 31\/}(2), 560--585.

\bibitem[\protect\citeauthoryear{Rigon}{Rigon}{2019}]{Rigon2019}
Rigon, T. (2019).
\newblock An enriched mixture model for functional clustering.
\newblock {\em arXiv:1907.02493\/}, 1--25.

\bibitem[\protect\citeauthoryear{Rousseau and Mengersen}{Rousseau and
  Mengersen}{2011}]{Rousseau2011}
Rousseau, J. and K.~Mengersen (2011).
\newblock {Asymptotic behaviour of the posterior distribution in overfitted
  mixture models}.
\newblock {\em Journal of the Royal Statistical Society. Series B: Statistical
  Methodology\/}~{\em 73\/}(5), 689--710.

\bibitem[\protect\citeauthoryear{Roy, Lum, Zeldow, Dworkin, Re, and
  Daniels}{Roy et~al.}{2018}]{Roy2018}
Roy, J., K.~J. Lum, B.~Zeldow, J.~D. Dworkin, V.~L. Re, and M.~J. Daniels
  (2018).
\newblock {Bayesian nonparametric generative models for causal inference with
  missing at random covariates}.
\newblock {\em Biometrics\/}~{\em 74\/}(4), 1193--1202.

\bibitem[\protect\citeauthoryear{Scarpa and Dunson}{Scarpa and
  Dunson}{2009}]{Scarpa2009}
Scarpa, B. and D.~B. Dunson (2009).
\newblock {Bayesian hierarchical functional data analysis via contaminated
  informative priors}.
\newblock {\em Biometrics\/}~{\em 65\/}(3), 772--780.

\bibitem[\protect\citeauthoryear{Scarpa and Dunson}{Scarpa and
  Dunson}{2014}]{Scarpa2014}
Scarpa, B. and D.~B. Dunson (2014).
\newblock {Enriched stick-breaking processes for functional data}.
\newblock {\em Journal of the American Statistical Association\/}~{\em
  109\/}(506), 647--660.

\bibitem[\protect\citeauthoryear{Sivaganesan, Laud, and
  M{\"{u}}ller}{Sivaganesan et~al.}{2011}]{Sivaganesan2011}
Sivaganesan, S., P.~W. Laud, and P.~M{\"{u}}ller (2011).
\newblock {A Bayesian subgroup analysis with a zero-enriched Polya Urn scheme}.
\newblock {\em Statistics in Medicine\/}~{\em 30\/}(4), 312--323.

\bibitem[\protect\citeauthoryear{ter Steege et~al.}{ter Steege
  et~al.}{2013}]{TerSteege2013}
ter Steege, H. et~al. (2013).
\newblock Hyperdominance in the amazonian tree flora.
\newblock {\em Science\/}~{\em 342\/}(6156), 1243092.

\bibitem[\protect\citeauthoryear{Wade, Dunson, Petrone, and Trippa}{Wade
  et~al.}{2014}]{Wade2014}
Wade, S., D.~B. Dunson, S.~Petrone, and L.~Trippa (2014).
\newblock {Improving prediction from Dirichlet Process mixtures via
  enrichment}.
\newblock {\em Journal of Machine Learning Research\/}~{\em 15}, 1041--1071.

\bibitem[\protect\citeauthoryear{Wade, Mongelluzzo, and Petrone}{Wade
  et~al.}{2011}]{Wade2011}
Wade, S., S.~Mongelluzzo, and S.~Petrone (2011).
\newblock {An enriched conjugate prior for Bayesian nonparametric inference}.
\newblock {\em Bayesian Analysis\/}~{\em 6\/}(3), 359--386.

\bibitem[\protect\citeauthoryear{Zeldow, Flory, Stephens-Shields, Raebel, and
  Roy}{Zeldow et~al.}{2021}]{Zeldow2021}
Zeldow, B., J.~Flory, A.~Stephens-Shields, M.~Raebel, and J.~A. Roy (2021).
\newblock {Functional clustering methods for longitudinal data with application
  to electronic health records}.
\newblock {\em Statistical Methods in Medical Research\/}~{\em 30\/}(3),
  655--670.

\bibitem[\protect\citeauthoryear{Zhang and Gao}{Zhang and
  Gao}{2020}]{Zhang2020}
Zhang, F. and C.~Gao (2020).
\newblock {Convergence rates of variational posterior distributions}.
\newblock {\em Annals of Statistics\/}~{\em 48\/}(4), 2180--2207.

\bibitem[\protect\citeauthoryear{Zito, Rigon, and Dunson}{Zito
  et~al.}{2022}]{Zito2022}
Zito, A., T.~Rigon, and D.~B. Dunson (2022).
\newblock Inferring taxonomic placement from dna barcoding allowing discovery
  of new taxa.
\newblock {\em arXiv:2201.09782\/}.

\end{thebibliography}

\end{document}